\documentclass[a4paper,11pt,english]{amsart}
\usepackage[applemac]{inputenc}
\usepackage[english]{babel}
\usepackage{amsfonts}
\usepackage{amsmath} 
\usepackage{amsthm}
\usepackage{hyperref} 
\usepackage{latexsym}
\usepackage{mathrsfs}
\usepackage{array}
\usepackage{amssymb}
\usepackage{enumerate}
\usepackage{smfthm}
\usepackage{graphicx}
\usepackage{color}
\usepackage{stmaryrd}
\usepackage{subfigure}
\newcommand{\R}{\mathbb  R}

\newcommand{\N}{\mathbb  N}

\renewcommand{\P}{\mathbb{P}}\newcommand{\dP}{\P}
\newcommand{\dE}{\mathbb{E}}

\newcommand{\eps}{\varepsilon}
\renewcommand{\epsilon}{\varepsilon}

\newcommand{\dN}{  \mathbb{N}   }
\newcommand{\Z}{  \mathbb{Z}   }

\renewcommand{\phi}{  \varphi  }

\newcommand{\IND}{1\!\!{\sf I}}

\numberwithin{equation}{section}
\theoremstyle{plain}

\def\beq{\begin{equation}}   \def\eeq{\end{equation}}
\def\bea{\begin{eqnarray}}  \def\eea{\end{eqnarray}}

\renewcommand{\theequation}{\thesection.\arabic{equation}}
\newcounter{hran} \renewcommand{\thehran}{\thesection.\arabic{hran}}

\def\bmini{\setcounter{hran}{\value{equation}}
    \refstepcounter{hran}\setcounter{equation}{0}
    \renewcommand{\theequation}{\thehran\alph{equation}}\begin{eqnarray}}

\def\bminiG#1{\setcounter{hran}{\value{equation}}
\refstepcounter{hran}\setcounter{equation}{-1}
\renewcommand{\theequation}{\thehran\alph{equation}}
\refstepcounter{equation}\label{#1}\begin{eqnarray}}

\author{Charles Bordenave}
\address{CNRS \& Universit\'e de Toulouse, Institut de Math\'ematiques de Toulouse, 118 route de Narbonne, 31062 Toulouse, France}
\email{charles.bordenave@math.univ-toulouse.fr}

\author{Pierre Germain}
\address{Courant Institute of Mathematical Sciences, 251 Mercer Street, New York 10012-1185 NY, USA
}
\email{pgermain@cims.nyu.edu}

\author{Thomas Trogdon}
\address{Courant Institute of Mathematical Sciences, 251 Mercer Street, New York 10012-1185 NY, USA
}
\email{trogdon@cims.nyu.edu}

\title[An extension of the Derrida-Lebowitz-Speer-Spohn equation ]{An extension of the Derrida-Lebowitz-Speer-Spohn equation}

\begin{document}

\begin{abstract}
We show how the derivation of the Derrida-Lebowitz-Speer-Spohn equation can be prolonged to obtain a new equation, generalizing the models obtained in the paper by these authors. We then investigate its properties from both an analytical and numerical perspective. Specifically, a numerical method is presented to approximate solutions of the prolonged equation.  Using this method, we investigate the relationship between the solutions of the prolonged equation and the Tracy--Widom GOE distribution. 
\end{abstract}

\thanks{P. Germain is partially supported by NSF grant DMS-1101269, a start-up grant from the Courant Institute, and a Sloan fellowship. Ch. Bordenave is partially supported by ANR-11-JS02-005-01.  T. Trogdon is partially supported by NSF grant DMS-1303018.}
\maketitle

 
\section{Introduction}

\subsection{The physics}

In \cite{DLSS}, Derrida, Lebowitz, Speer and Spohn proposed a simplified model to describe the low temperature Glauber dynamics of the North-East model in the presence of two phases with an anchored interface.  It can be described as a Markov process $\eta(t)$ on $\{-1,1\}^{n}$. The model has two parameters, $\lambda_+,\lambda_- >0$. Informally, at time $t \geq 0$, each site $x \in \{1, \ldots, n\}$ has an independent alarm clock which rings after an exponential random variable with mean $1/\lambda_{\eta_x(t)}$. When the first alarm rings, say at site $x$, we exchange the values of $\eta_x(t)$ and $\eta_{y}(t)$ where, if it exists, $y$ is the minimum of all $x < z \leq n$ such that $\eta_{z} (t) = - \eta_{x}(t)$. If for all $z > x$, $\eta_z(t) = \eta_x(t)$ we nevertheless invert the value of $\eta_x(t)$. More formally, for $ \eta \in \{-1,1\}^n$, the exchange rate $c_{x,y}(\eta)$ between two sites $1 \leq x < y \leq n$ is defined as
$$
c_{x,y}(\eta) =  \lambda_+  \frac{ (1- \eta_y)} 2  \prod_{x \leq z < y}  \frac{ (1 + \eta_x)}{2} +  \lambda_-  \frac{ (1+ \eta_y)} 2  \prod_{x \leq z < y}  \frac{ (1 - \eta_x)}{2},
$$
and the flip rate $c_x (\eta)$ is defined by 
$$
c_{x}(\eta) =  \lambda_+   \prod_{x \leq z  \leq n }  \frac{ (1 + \eta_x)}{2} +  \lambda_-    \prod_{x \leq z \leq n}  \frac{ (1 - \eta_x)}{2}.
$$
Then, we consider the Markov process $\eta(t) \in \{-1,1\}^{n}$ which exchanges and flips the values of its coordinates with the above transition rates. This process has the beautiful property that $(\eta_x(t))_{1 \leq x \leq k}$ its restriction  to $\{1, \ldots, k\}$, with $k < n$, again follows the same dynamics. From the Kolmogorov extension theorem, we can define a Markov process $\eta(t)$ on $\{-1,1\}^{\dN}$ whose restriction to $(\eta_{x}(t))_{1 \leq x \leq n}$ is the above Markov process.  It is not difficult to check that, for each $n\geq 1$ the processes $(\eta_x (t))_{1 \leq k \leq n}$  is an irreducible Markov process. It has a unique invariant measure. In particular the whole Markov process on $\{-1,1\}^{\dN}$ admits a unique invariant measure which we will denote by $\dP$. 

We can alternatively interpret the process as an interacting particle system. For $\eta \in \{-1,1\}^{\dN}$, we say that sites such that $\eta_x = 1$ are occupied by a particle and sites such that $\eta_x = -1$ are empty. A site can be occupied by at most one particle. We denote the position of the particles by $1 \leq X_1 < X_2 < \cdots$, i.e. for integer $k \geq 1$, $\eta_{X_k} = 1$ and $\eta_{X_{k-1} + \ell} = -1$ for $1 \leq \ell < X_{k} - X_{k-1}$ (with the convention $X_0 = 0$). Then, the Markov process for the particles $X(t) = ( X_1 (t), X_2 (t), \ldots)$ is described as follows. For each $x \in \{ X_{k-1} (t) +1, \cdots,  X_k(t)-1 \}$, at rate $\lambda_-$, the $k$-th particle jumps from $X_{k} (t)$ to $x$. For each $k$, at rate $\lambda_+$, the $k$-th particle jumps by one step on its right: it jumps to $X_{k}(t) +1$ and pushes all its right neighbors by $1$ if they prevent it from jumping to the right. This equivalent description of the process shows its similarity with simple exclusion processes which have been studied extensively in integrable probability, see notably \cite{MR2438811,CP2013,MR3018876} and for general overviews \cite{MR2827973,dotsenko,BP2013}. Other closely related models of spin exchanges are studied in \cite{MR1424235,ASST2014}. It seems however that despite the models studied in the aforementioned articles, the Markov process $X(t)$ has no known closed form formula for its marginal at time $t$. Nevertheless, the process has again the restriction property that the process of the first $k$-th particle $(X_1(t), \cdots, X_k(t))$ follows the same Markovian dynamics. Using this property, it is again not difficult to check that the Markov process $X(t)$ admits a unique invariant measure on $\dN^\dN$. 

\subsection{The magnetization and its asymptotic behaviour}

In \cite{DLSS}, the authors are mainly interested by the stationary magnetization. It is the random variable
$$
M_n = \sum_{x=1}^n \eta_x, 
$$
where $\eta$ has the invariant distribution of the Markov process.  The variable can also be easily deduced from the particle system $X$ in stationary regime
$$
M_n \geq m \quad  \Leftrightarrow \quad X_{\frac{m + n}{2}} \leq n. 
$$

For the remainder of this paper, we set
$$
\mu = \frac{\sqrt{\lambda_-} - \sqrt{\lambda_+}}{\sqrt{\lambda_-} + \sqrt{\lambda_+}}
\quad \mbox{and} \quad C = \frac{\sqrt{\lambda_- \lambda_+}}{(\sqrt{\lambda_-} + \sqrt{\lambda_+})^2}
$$
(observe that $0 \leq C \leq \frac{1}{4}$ and $-1 \leq \mu \leq 1$).
The case $\mu = 0$ is called the unbiased (or symmetric) case. The case $\mu \ne 0$, the biased (or asymmetric) case. In \cite{DLSS}, based on a non-rigorous approximation, the authors conjecture that if $\mu = 0$, a central limit theorem holds for $M_n$, for any $x \in \R$,
\begin{equation}\label{EW}
\lim_{n \to \infty} \dP  \left( \frac{ M_n }{ (  ( 3/2 )   n )^{1/4}} \geq  x \right)   =\frac 1 {\sqrt{2 \pi}}  \int_ x ^\infty e^{-\frac {s^2}{ 2}} ds. 
\end{equation}
In the unbiased case,  $\mu \ne 0$, \cite{DLSS} conjectures that a different scaling and weak limit appear. Namely, for all $x \in \R$,
\begin{equation}\label{KPZ}
\lim_{n \to \infty} \dP  \left( \frac{ M_n - \mu n  }{ ( \mu C n ) ^{1/3}} \geq  x \right)  = Q ([x, \infty) ), 
\end{equation}
for some probability measure $Q$ on $\R$ independent of $(\lambda_+,\lambda_-)$.  In the unbiased case, the scaling of the variance as $\sqrt n$ suggests that the unbiased process falls into the universality class of Edwards-Wilkinson. In the biased case, the scaling $n^{1/3}$ suggests the KPZ universality class, for a recent survey on the latter see \cite{MR2930377}. 

\subsection{A new PDE governing the equilibrium measure}

\begin{figure}[tb]
\includegraphics[width=.5\linewidth]{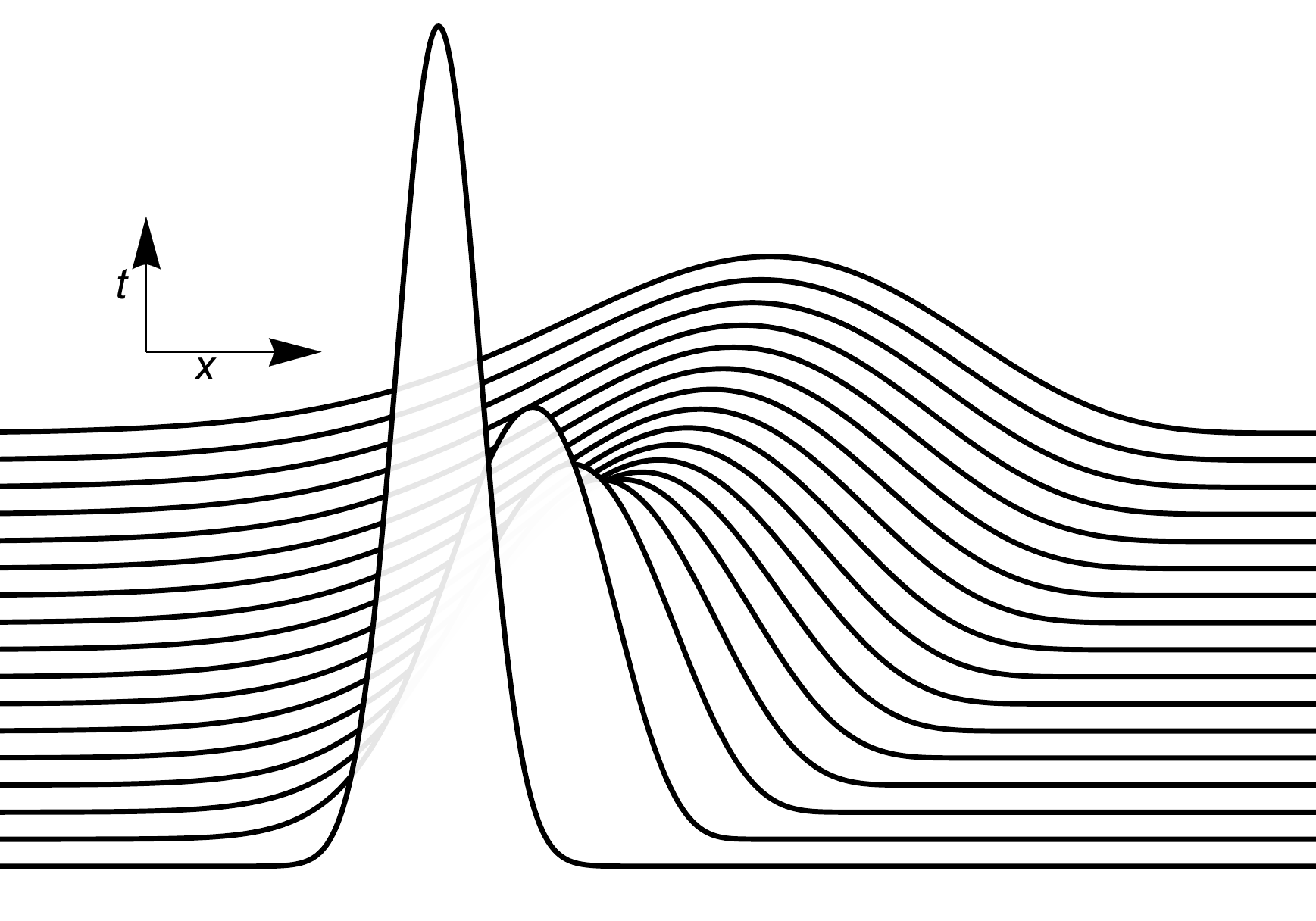}
\caption{\label{Gaussian} The evolution of a Gaussian initial condition under the equation \eqref{eqw} when $\mu > 0$.}
\end{figure}

In \cite{DLSS}, the authors derive non-rigorously a PDE associated to the rescaled marginal of $M_n$. In the present paper, we will revisit the computation of \cite{DLSS} and obtain a generalization of their PDE. To be more precise, we introduce the functions 
$$
\left\{ 
\begin{array}{l} 
H_n(m) =  \dE [ \eta_{n+1} | M_n = m] \\
U^{\pm}_n(m) =\lambda_{\pm} \dE [ K_n^{\pm} \IND ( M_n = m)] \\
W_n(m) =  \dP ( M_n = m),
\end{array}
\right.
$$
 where $\dE$ denotes the expectation with respect to the stationary measure $\dP$, $\IND$ is the indicator function and $K_n^{\pm}$ is the number of successive left neighbors of site $n$ (including $n$) such that $\eta_x = \pm 1$.  The stationarity of the process implies that $U^+_n(m) = U^-_n(m)$ (\cite[Eqn (4.18)]{DLSS}). We set their common value to be $U(m)$. 

We adopt the ansatz
$$
\left\{ 
\begin{array}{l} 
H_n(m) = \mu + h(\epsilon^3 n,\epsilon (m - \mu n)) \\
U_n(m) = u(\epsilon^3 n,\epsilon(m-\mu n))\\
W_n(m) = w(\epsilon^3 n,\epsilon(m-\mu n))
\end{array}
\right.
$$

Observe that the scaling between $n$ and $m$ is consistent with \eqref{KPZ}.  Under a non-rigorous approximation, we show that under this scaling, as $\epsilon \to 0$, $w$ is governed by the following partial differential equation:
\begin{equation}
\label{eqw}
\boxed{\partial_t w - \mu C \left( \frac{4}{3} \partial_x^3 w - \partial_x \left( \frac{(\partial_x w)^2}{w} \right) \right) = \epsilon (2C^2 - 2C)\left( \partial_x^4 w - \partial_x^2 \left( \frac{(\partial_x w)^2}{w} \right) \right) }
\end{equation}
(notice that $0 \leq 2C - 2C^2 \leq \frac{1}{4}$),
which can also be written
$$
\partial_t w - \mu C \partial_x \left( w^{3/4} \partial_x^2 w^{1/4} \right) =
\epsilon (2C^2 - 2C) \partial_x^2 \left( w \partial_x^2 \log w\right).
$$
For $\mu = 0$, the leading term vanish and the presence of $\eps$ agrees with \eqref{EW}. The equation for $v$ such that $w=v^2$ is somewhat simpler: it reads
\begin{equation}
\label{eqv}
\boxed{\partial_t v - \frac{4}{3} \mu C \partial_x^3 v = \epsilon(2C^2-2C) \left( \partial_x^4 v - \frac{(\partial_x^2 v)^2}{v} \right)}
\end{equation}
or equivalently
$$
\partial_t v - \frac{4}{3} \mu C \partial_x^3 v = \epsilon(2C^2-2C)  \frac{1}{v} \partial_x \left( v^2 \partial_x \left( \frac{\partial_x^2 v}{v} \right) \right).
$$
The equations~\eqref{eqw} and \eqref{eqv} generalize the PDEs found in~\cite{DLSS}, which correspond to the cases $\mu = 0$ and $C = \frac{1}{4}$, or $\mu \neq 0$ and $\epsilon = 0$.

The equation obtained in the unbiased case $\mu =0$, $C = \frac{1}{4}$ has been the subject of intensive research in the PDE community, see~\cite{BLS,CCT,DGJ,GST,JM,JM2,JM3,MMS} and the references therein; we will come back to the results obtained in these papers in Section~\ref{sectionPDE}.

In the biased case $\mu \neq 0$, the equation derived in the present paper, namely~\eqref{eqv}, corresponds to adding a right-hand side to the equation derived in~\cite{DLSS}. This seems to reflect an important correction, since it has a dissipative behaviour, thus potentially giving (up to rescaling!) a trend towards a universal profile as $t \to \infty$. We could not identify this profile in the case where the equation is set in $\mathbb{R}$, which is the most interesting one, but it should be related to the crucial question of the asymptotic invariant measure of the random process under consideration.

\subsection{A numerical investigation}

We also perform a detailed numerical investigation of the solutions of \eqref{eqw}.  See Figure~\ref{Gaussian} for an example solution.  The numerical solution of \eqref{eqw} is difficult from two points of view.
\begin{enumerate}
\item The nonlinearities in \eqref{eqw} are singular so that the equation must be rewritten for numerical purposes.
\item If $w > 0$ initially, then we argue below that $w$ should be positive for all time.  The third-order linear term in \eqref{eqw} works to make the function vanish while the nonlinearities prevent this.  There is a strong, non-trivial coupling between these terms and split-step methods, which are standard in the numerical solution of nonlinear dispersive equations, cannot be used.
\end{enumerate}
We are able to overcome these issues and simulate \eqref{eqw} for moderate times with a stable, highly-accurate pseudospectral scheme.  We provide evidence that
\begin{align}\label{intro-limit}
w(t,x) \sim \frac{1}{t^{1/3}} f \left( \frac{ x}{t^{1/3}} \right),
\end{align}
for some, yet unknown, function $f$.  The authors in \cite{Toom} use Monte Carlo simulations to approximate the asymptotic equilibrium measure for the random process we consider.  They show that it is approximated well with the Tracy--Widom ($\beta =1$) GOE distribution \cite{Tracy1996}. Because we expect the long-time behavior of $w$ to be related to the asymptotic invariant measure, we compare $w(t,x)$ with the Tracy--Widom GOE distribution.  After normalization, the distance of the solution from this Tracy--Widom distribution is found to be on the order of that  in \cite{Toom}.  Due to the high accuracy of our numerical method, and the fact that $f$ in \eqref{intro-limit} is, in our computations, distinct from the Tracy--Widom GOE density, we raise the question of whether the asymptotic equilibrium measure for the random process could be something other than the Tracy--Widom GOE distribution.

\subsection{An instructive analogy: the sum of independent random variables}

In order to better understand the computations which will follow, let us start with the simple example of independent and identically distributed (iid) variables. Assume that $(\sigma_k)_{k \geq 1}$ is a sequence of iid variables on $\{-1,1\}$ with $\nu = \dE \sigma_k$, i.e. $\dP ( \sigma_k = 1) = 1 - \dP ( \sigma_k = -1) = (1 + \nu  ) /2$.  We set $$S_n = \sum_{k=1} ^ n \sigma_k$$ 
and $P_n ( m ) = \dP ( S_n = m )$. We have the recursion 
\begin{equation}\label{heat}
P_{n+1} (m) = \frac{ 1 + \nu}{2} P_n (m-1) +  \frac{ 1 - \nu}{2} P_n (m+1). 
\end{equation}
The convergence of a properly rescaled version of $P_n$ to the heat equation could be obtained heuristically as follows. For any $\eps >0$, we may define a function $p_\eps(t,x)$ such that $P_n (m) = p_\eps ( \eps^2 n , \eps ( m - \nu n ) )$. We 
may rewrite \eqref{heat} as, if $t = \eps^2 n $ and $x = \eps ( m - \nu n)$, 
$$
p_\eps ( t  + \eps^2, x - \nu \eps ) =  \frac{ 1 + \nu}{2} p_\eps ( t , x  - \eps)  +  \frac{ 1 - \nu}{2} p_\eps (t , x + \eps). 
$$
We now let $\eps \to 0$.  The central limit theorem implies notably that $p_\eps /  \eps $ converges to a probability density  function $p$. We expand in powers of $\eps$ the above identity. The first non-zero term is in $\eps^2$, it gives the PDE
$$
\partial_t p  =  \theta \partial^2_x p. 
$$
with $\theta = ( 1 - \frac {\nu^2}{2} )$. We recognize the heat equation in one dimension. Also, for any $\eps, \eps' >0$, we have that $P_n (m) = p_\eps ( \eps^2 n , \eps ( m - \nu n ) ) = p_{\eps'}( \eps'^2 n , \eps' ( m - \nu n ) ) $. Hence, for any $s >0$, if we consider the case $\eps' = s \eps$ and $\eps \to 0$, we deduce that the probabilistically relevant solution of the PDE should also satisfy, $p ( t, x) = s p ( s^2 t , s x)$.  In other words, they should be of the form 
$$
p(t,x) = g ( x / \sqrt t) / \sqrt t, 
$$
for a probability density function $g$. It follows that $g$ satisfies an ODE which we can of course explicitly solve in this simple case and retrieve the Gaussian density. As in \cite{DLSS}, for the DLSS Markov process, we will follow a similar strategy.

\subsection{Plan of the paper}  The non-rigorous derivation of \eqref{eqw} and \eqref{eqv} is presented in Section~\ref{sectionderivation}, while some properties of these equations are analyzed formally in Section~\ref{sectionPDE} . In Section~\ref{sectioninvariant}, we discuss the invariant measure for the first particles of the above stochastic process.  Finally, we present a method for the numerical solution of \eqref{eqw} and a detailed analysis of the approximate solutions in Section~\ref{section:numerics}.

\section{Derivation of the extended DLSS equation}

\label{sectionderivation}

\subsection{Outline of the derivation}

We will use the equations
\begin{align}
\label{pingouin1}
& W_{n+1}(m) = \frac{1}{2} (1+H_n(m-1))W_n(m-1) + \frac{1}{2} (1 - H_n(m+1))W_n(m+1) \\
\label{pingouin2}
& H_n(m) = \frac{U_n(m+1) - U_n(m-1) + (\lambda_- - \lambda_+) W_n(m)}{U_n(m+1) + (\lambda_- + \lambda_+) W_n(m) + U_n(m-1)} \\
\label{pingouin3}
& U_{n+1}(m) = \frac{(\lambda_- W_n(m) + U_n(m+1) (\lambda_+ W_n(m) + U_n(m-1))}{U_n(m+1) + (\lambda_- + \lambda_+) W_n(m) + U_n(m-1)}.
\end{align}
which appear in~\cite{DLSS} as (6.4), (6.9), and (6.10) respectively. These equations rely on the simplifying approximation that, given $M_n$, $\eta_{n+1}$ is (approximately) independent of $K^{\pm}_n$. This could be justified heuristically by  observing that the DLSS Markov process now defined on $\{-1,1\}^{\Z/n\Z}$ instead of $\{-1,1\}^\N$ preserves the number of $+$ and $-$ sites and, given the number of $+$ and $-$ sites,  the invariant probability measure is the uniform measure (see \cite{DLSS}).  Hence, we may expect that when $n$ and $m$ are large, $\dP ( \eta_{n-\ell} = a_\ell , -k \leq \ell \leq k | M_n = m )$ could be approximated by $\prod_{-k \leq \ell \leq k} \dP ( \eta_{n - \ell} = a_ \ell | M_n = m)$. 

Our plan is now as follows
\begin{enumerate}
\item Expand $u$ in powers of $\epsilon$, with coefficients depending on $w$ (Section~\ref{merle1}).
\item Expand $wh$ in powers of $\epsilon$, with coefficients depending on $w$ (Section~\ref{merle2}).
\item Find the equation satisfied by $w$ (Section~\ref{merle3}).
\end{enumerate}

\subsection{Expansion of $u$ in $\epsilon$}

\label{merle1}

We start with the ansatz
$$
u = \alpha + \epsilon \beta + \epsilon^2 \gamma + \epsilon^3 \delta
$$
and aim at determining $\alpha$, $\beta$, $\gamma$ and $\delta$ as functions of $w$.
First, expanding the left-hand side of \eqref{pingouin3} to order 3 gives
\begin{equation}
\label{chardonneret}
\begin{split}
LHS\eqref{pingouin3} = \alpha + \epsilon(\beta - \mu & \partial_x \alpha) + \epsilon^2 \left( \gamma - \mu \partial_x \beta + \frac{1}{2} \mu^2 \partial_x^2 \alpha \right)\\
&  + \epsilon^3 \left( \delta + \partial_t \alpha - \mu \partial_x \gamma + \frac{1}{2} \mu^2 \partial_x^2 \beta - \frac{1}{6} \mu^3 \partial_x^3 \alpha \right) + O(\epsilon^4)
\end{split}
\end{equation}
while expanding the right-hand side of \eqref{pingouin3} to order 1 yields
\begin{equation*}
\begin{split}
RHS(\ref{pingouin3}) & = \frac{(\lambda_- w + \alpha + \epsilon \beta + \epsilon \partial_x \alpha)(\lambda_+ w + \alpha + \epsilon \beta - \epsilon \partial_x \alpha)}{2 \alpha + 2 \epsilon \beta + (\lambda_+ + \lambda_-) w} + O(\epsilon^2) \\
& = \frac{(\lambda_- w + \alpha)(\lambda_+ w + \alpha)}{2 \alpha + (\lambda_+ + \lambda_-)w} \\
& \qquad + \epsilon \left[  \frac{\beta(2\alpha + (\lambda_+ + \lambda_-)w) + \partial_x \alpha (\lambda_+ - \lambda_-)w}{2 \alpha + (\lambda_+ + \lambda_-)w} - \frac{2\beta (\lambda_+ w + \alpha) (\lambda_- w + \alpha)}{(2 \alpha + (\lambda_+ + \lambda_-)w)^2} \right] + O(\epsilon^2).
\end{split}
\end{equation*}
Identifying terms of order 0 and 1 in $\epsilon$ in the left- and right-hand sides of \eqref{pingouin3} leads to
$$
\alpha = \sqrt{\lambda_+ \lambda_-} w \qquad \mbox{and} \qquad \beta=0.
$$
Next, expand the right-hand side of (\ref{pingouin3}) to order 3 in $\epsilon$, taking advantage of the fact that $\beta=0$. This gives
\begin{equation*}
\begin{split}
RHS(\ref{pingouin3}) & = \left( \lambda_- w + \alpha + \epsilon^2 \gamma + \epsilon^3 \delta + \epsilon \partial_x \alpha + \epsilon^3 \partial_x \gamma + \frac{\epsilon^2}{2}\partial_x^2 \alpha + \frac{\epsilon^3}{6}\partial_x^3 \alpha \right) \\
& \qquad \qquad \times \left( \lambda_+ w + \alpha + \epsilon^2 \gamma + \epsilon^3 \delta - \epsilon \partial_x \alpha - \epsilon^3 \partial_x \gamma + \frac{\epsilon^2}{2} \partial_x^2 \alpha - \frac{\epsilon^3}{6}\partial_x^3 \alpha \right) \\
& \qquad \qquad \times \frac{1}{2\alpha + 2\epsilon^2 \gamma + 2 \epsilon^3 \delta + \epsilon^2 \partial_x^2 \alpha + (\lambda_+ + \lambda_-)w}+  O(\epsilon^4)\\
& =  A + B \epsilon + C \epsilon^2 + D \epsilon^3 + O (\epsilon^4)
\end{split}
\end{equation*}
where $A$ and $B$ have already been determined, and
\begin{equation*}
\begin{split}
& C = \frac{-(\partial_x \alpha)^2 + \gamma ( (\lambda_+ + \lambda_-) w + 2\alpha) + \partial_x^2 \alpha(\alpha+ \frac{1}{2}(\lambda_+ + \lambda_-) w)}{2 \alpha + (\lambda_+ + \lambda_-) w} - \frac{(2 \gamma + \partial_x^2 \alpha)(\lambda_+ w + \alpha)(\lambda_- w + \alpha)}{(2 \alpha + (\lambda_+ + \lambda_-) w)^2} \\
& D = \delta - \frac{2\delta(\lambda_- w + \alpha)(\lambda_+ w + \alpha)}{(2 \alpha + (\lambda_+ + \lambda_-) w)^2} - \frac{\partial_x \alpha (\lambda_+ - \lambda_-) w (2 \gamma + \partial_x^2 \alpha)}{(2 \alpha + (\lambda_+ + \lambda_-) w)^2} + \frac{(\partial_x \gamma + \frac{1}{6} \partial_x^3 \alpha)(\lambda_+ - \lambda_-) w}{2 \alpha + (\lambda_+ + \lambda_-) w}.
\end{split}
\end{equation*}
Using the equality $\alpha = \sqrt{\lambda_+ \lambda_-} w$ as well as the definitions of $C$ and $\mu$ leads to the more simple formulas
\begin{equation*}
\begin{split}
& C = \sqrt{\lambda_+ \lambda_-} \left(\frac{1}{2}-C\right) \partial_x^2 w -\sqrt{\lambda_+ \lambda_-} C \frac{(\partial_x w)^2}{w} + (1-2C) \gamma \\
& D = (1-2C)\delta + \mu C \sqrt{\lambda_+ \lambda_-} \frac{\partial_x w}{w} \left( 2 \partial_x^2 w - \frac{(\partial_x w)^2}{w} \right) - \mu \partial_x \gamma - \frac{\mu}{6}\partial_x^3 \alpha
\end{split}
\end{equation*}
and identifying the terms of order 2 in $\epsilon$ in the left- and right-hand sides of (\ref{pingouin3}) gives, respectively,
\begin{equation*}
\begin{split}
& \gamma = \frac{\sqrt{\lambda_+ \lambda_-}}{2} \left( \partial_x^2 w - \frac{(\partial_x w)^2}{w} \right) \\
& \delta = \sqrt{\lambda_+ \lambda_-} \left(-\frac{1}{2C} \partial_t w - \frac{\mu}{3} \partial_x^3 w + \mu \frac{\partial_x w \partial_x^2 w}{w} - \frac{\mu}{2} \frac{(\partial_x w)^3}{w^2} \right).
\end{split}
\end{equation*}

\subsection{Expansion of $wh$ in $\epsilon$}

\label{merle2}

Expanding the right-hand side of \eqref{pingouin2} to order 3 in $\epsilon$ gives
$$
\mu + h = \frac{(\lambda_- - \lambda_+) w + 2\epsilon \partial_x \alpha + 2 \epsilon^3 \partial_x \gamma + \frac{1}{3} \epsilon^3 \partial_x^3 \alpha}{2 \alpha + 2 \epsilon^2 \gamma + 2 \epsilon^3 \delta + \epsilon^2 \partial_x^2 \alpha + (\lambda_+ + \lambda_-) w}
+ O(\epsilon^4) $$
or, after replacing $\alpha$, $\gamma$ and $\delta$ by the formulas derived above,
\begin{equation}
\label{cormoran}
\begin{split}
& wh = \epsilon 2 C \partial_x w + \epsilon^2  \mu C \left( -2 \partial_x^2 w + \frac{(\partial_x w)^2}{w} \right) \\
& \qquad \qquad + \epsilon^3 \left( \mu \partial_t w + C \left( \frac{4}{3}+\frac{2\mu^2}{3}\right) \partial_x^3 w - C(1+\mu^2+2C)\partial_x \left( \frac{(\partial_x w)^2}{w} \right)\right) +O(\epsilon^4) .
\end{split}
\end{equation}

\subsection{Equation satisfied by $w$}

\label{merle3}

The left-hand side of \eqref{pingouin1} reads, to order 4 in $\epsilon$,
$$
LHS\eqref{pingouin1} = w + \epsilon^3 \partial_t w - \mu \epsilon \partial_x w + \frac{1}{2} \mu^2 \epsilon^2 \partial_x^2 w - \frac{1}{6} \mu^3 \epsilon^3 \partial_x^3 w + \frac{1}{24} \mu^4 \epsilon^4 \partial_x^4 w - \mu \epsilon^4 \partial_t \partial_x w + O(\epsilon^5),
$$
while the right-hand side of \eqref{pingouin1} can be expanded as
\begin{equation*}
\begin{split}
RHS\eqref{pingouin1}& = w + \frac{\epsilon^2}{2} \partial_x^2 w + \frac{\epsilon^4}{24} \partial_x^4 w - \mu \epsilon \partial_x w - \mu \frac{\epsilon^3}{6} \partial_x^3 w - \epsilon \partial_x(hw) - \frac{\epsilon^3}{6} \partial_x^3 (hw) + O(\epsilon^5), \\
\end{split}
\end{equation*}
which, with the help of~\eqref{cormoran}, gives
\begin{equation*}
\begin{split}
RHS\eqref{pingouin1} & = w - \epsilon \mu \partial_x w + \epsilon^2 \left(\frac{1}{2}-2C\right) \partial_x^2 w
+ \epsilon^3\left( \left(-\frac{\mu}{6} + 2 \mu C \right) \partial_x^3 w - \mu C \partial_x \left(\frac{(\partial_x w)^2}{w} \right) \right) \\
&  \;\;\;\; + \epsilon^4 \left(\left((-\frac{2}{3}\mu^2 C - \frac{5C}{3} + \frac{1}{24} \right) \partial_x^4 w - \mu \partial_t \partial_x w + C (1+\mu^2 + 2C) \partial_x \left( \frac{(\partial_x w)^2}{w} \right)\right) + O(\epsilon^5).
\end{split}
\end{equation*}
Equating terms of order 4 and 5 on the left- and right-hand sides of \eqref{pingouin1} yields
$$
\partial_t w - \frac{4}{3} \mu C \partial_x^3 w + \mu C \partial_x \left( \frac{\partial_x w)^2}{w} \right) = \epsilon (2C^2 -2C) \left(\partial_x^4 w - \partial_x^2 \left( \frac{(\partial_x w)^2}{w} \right) \right),
$$ 
which is the desired result.

\section{A few properties of the equation}

\label{sectionPDE}

We examine in this section some of the properties of~(\ref{eqw}) and~(\ref{eqv}); we stay at a formal level and do not try to give rigorous proofs. In order to alleviate the notations, we denote in the following 
$$
K = \frac{4}{3} \mu C \qquad \mbox{and} \qquad L = \epsilon(2C - 2C^2) 
$$
(notice that $L \geq 0$). We let the independent variables  $(t,x)$ range over $\mathbb{R_+} \times \mathbb{T}$ or $\mathbb{R_+} \times \mathbb{R}$, the second case being physically more relevant.

The equation on $w$ (taking values in $\mathbb{R}_+$) reads now
\begin{equation}
\label{cormoran1}
\partial_t w - K \left( \partial_x^3 w - \frac 3 4 \partial_x \left( \frac{(\partial_x w)^2}{w} \right) \right) = - L \left( \partial_x^4 w - \partial_x^2 \left( \frac{(\partial_x w)^2}{w} \right)\right)
\end{equation}
while the equation on $v$ (taking values in $\mathbb{R}$) is given by 
\begin{equation}
\label{cormoran2}
\partial_t v - K \partial_x^3 v = - L \left( \partial_x^4 v - \frac{(\partial_x^2 v)^2}{v}  \right).
\end{equation}

\subsection{Questions of sign} It is a delicate question to understand how the equations~(\ref{cormoran1}) and~(\ref{cormoran2}) are exactly related, the difficulty arising at points where $v$ or $w$ vanish. Assuming that $v$ is smooth and setting $w = v^2$, a small computation gives that
$$
\partial_t w - K \left( \partial_x^3 w - \frac{3}{4} \partial_x \left( \frac{(\partial_x w)^2}{w} \right) \right) + L \left( \partial_x^4 w - \partial_x^2 \left( \frac{(\partial_x w)^2}{w} \right)\right) = 2 v \left[ \partial_t v - K \partial_x^3 v + L \left( \partial_x^4 v - \frac{(\partial_x^2 v)^2}{v}  \right) \right],
$$
so that $w$ solves~(\ref{cormoran1}) in a weak sense if $v$ solves (\ref{cormoran2}). Notice that $w$ defined by $w=v^2$ is automatically non-negative. On the other hand, it is not clear whether one can always lift ~(\ref{cormoran1}) to (\ref{cormoran2}).

\medskip \noindent \underline{Non-vanishing solutions.} If $v \geq 0$ and $L>0$, a heuristic argument (which was pointed out to us by Percy Deift) implies that $v$ should keep a constant sign. Indeed, assume that $v$ vanishes at a later time, say at $(t_0,x_0)$. Generically, it happens in such a way that $\partial_x^2 v(t_0,x_0) >0$. But then, as $t \to t_0$, $\frac{(\partial_x^2 v)^2}{v} \to \infty$, which implies $\partial_t v(t_0,x_0) = \infty$, which contradicts the vanishing of $v$ at $(t_0,x_0)$.

\medskip \noindent \underline{Vanishing solutions.} If $L = 0$, it is well-known that solutions to~\eqref{cormoran2} develop zeros as $t \to \infty$, even if they are not present at the initial time. If $v(t=0)$ has zeros, they are, at least locally in time, conserved: see the exact solutions below for some examples. Denoting $X(t)$ for one of the zeros of $v$, one can locally expand $v$ in powers of $(x-X)^2$; this shows easily that $\partial_x^2 v(t,X(t)) = 0$. At the level of $w$, it should be expected, as in the case $K=0$, that zeros are unstable.
 
\subsection{Symmetries}

Space or time translations of course leave the equation invariant. A more interesting symmetry is given by
$$
w \mapsto \lambda w \qquad \mbox{and} \qquad v \mapsto \lambda v
$$
(where $\lambda$ is non-negative). However, the probabilistic interpretation of the equation requires that $w$ be the density of a probability measure, making its multiplication by a non-negative number physically irrelevant.

For $K=0$ or $L=0$, the equation has a scaling symmetry ($v\mapsto v(\lambda^3 t,\lambda x)$ and $v \mapsto v(\lambda^4 t, \lambda x)$ respectively), which is lost for general $K$ and $L$. The gradient flow structure noticed and exploited in~\cite{GST,MMS} for $K = 0$ is also lost if $K \neq 0$.

\subsection{Lyapunov functions} On the one hand, it was first noticed in~\cite{BLS} that quantities of the type $\int |w|^\alpha$ or $\int |(w^\beta)_x|^2$ are monotonic for solutions of (\ref{cormoran1}) if $K=0$. The range of $\beta$ was later extended in~\cite{JM}. On the other hand, (\ref{cormoran1}) is simply Airy's equation if $L=0$, for which conserved quantities are well-known: $\int v$ and all the $L^2$-based Sobolev norms $\int |\partial_x^s v|^2$. It is not surprising that Lyapunov functions for the general case $K,L \neq 0$ correspond to these quantities which are invariant or monotonic both if $K=0$ and $L=0$:
\begin{itemize}
\item The "mass" $\displaystyle \int w \,dx = \int v^2\,dx$ of $w$ is conserved: $\displaystyle \frac{d}{dt} \int w\,dx = 0$ (since $w$ models a density of probability, the physical interpretation is clear).
\item The "momentum" $\displaystyle \int \sqrt{w} \,dx = \int v\,dx$ of $v$ is increasing: $\displaystyle \frac{d}{dt} \int v \,dx= L \int \frac{(\partial_x^2 v)^2}{v}\,dx$.
\item The "Fisher information" $\displaystyle \int \frac{(\partial_x w)^2}{w}\,dx = \int (\partial_x v)^2\,dx$ of $w$ is decreasing: 
$$\displaystyle \frac{d}{dt} \int v_x^2\,dx = L \int \partial_x^2 v \left( \partial_x^4 v - \frac{(\partial_x^2 v)^2}{v} \right) \,dx = - L \int \left( \partial_x^3 v - \frac{\partial_x^2 v \partial_x v}{v} \right)^2\,dx.$$
\item Finally, the "mean" $\displaystyle \int w x\,dx = \int v^2 x\,dx$ varies according to
$$
\frac{d}{dt} \int v^2 x \,dx = \frac{3}{2} K \int (\partial_x v)^2\,dx
$$
(so that $\frac{d^2}{dt^2} \int v^2 x \,dx$ has the same sign as $-K$, indicating the direction in which $v$ has a tendency to drift).
\end{itemize}

\subsection{Exact solutions} For $K=0$, examples of exact solutions were first given in~\cite{BLS}; we show how some of these examples can be extended to the case $K \neq 0$, providing some insight into the dynamics of this new equation.
\begin{itemize}
\item First, it is immediate to check that $v = \sin(x - Kt)$ is an exact traveling wave solution, which becomes stationary if $K = 0$.
\item Similarly for $v = \sinh(x + Kt)$.
\item Next, it was already noticed that the Airy functions $\operatorname{Ai}(x) = \int_0^\infty \cos(tx + t^3/3)\,dt$ gives, if $K=0$, the traveling wave $v = \operatorname{Ai}(x - 2 Lt)$. It is also well-known that $\frac{1}{t^{1/3}} \operatorname{Ai} \left( \frac{x}{(-3Kt)^{1/3}} \right)$ is a solution of the Airy equation obtained if $L=0$. For $K,L \neq 0$, we were able to find an exact solution based on the Airy function:
$$v(t,x) = \frac{1}{t^{1/3}} \operatorname{Ai} \left( \frac{x + \frac{2L}{3K} \log t}{(-3Kt)^{1/3}} \right)$$
(this formula can be checked directly using the fact that the Airy function $\operatorname{Ai}$ solves the ODE $y''=xy$).
\item This remains true if $\operatorname{Ai}$ is replaced by the Airy function of the second kind often denoted $\operatorname{Bi}$.
\end{itemize}
Though these explicit solutions certainly help understanding better the equation, it is not clear how much they say about its large time behavior in the case of finite mass, which is physically relevant.

Indeed, if one thinks of the setting where $x \in \mathbb{T}$, the only acceptable solution in the above list is the periodic one $v = \sin(x + Kt)$. However, it is known to be unstable, at least in the case $K=0$, see~\cite{JM2}. As for the setting where $x \in \mathbb{R}$, the above solutions all have infinite mass, even though the one based on the Airy function decays at infinity.

\subsection{Asymptotic behaviour}

\subsubsection{The case $x \in \mathbb{T}$} Without loss of generality, we assume here that $\int w =1$. For $L=0$, there is no trend to equilibrium, and $w$ oscillates indefinitely. For $K=0$, it was proved in~\cite{CCT,DGJ} that $w$ converges exponentially fast to the constant $w \equiv 1$ (see also~\cite{GST} for a much more general framework). This remains true for $K,L \neq 0$: we claim that there exists $\mu>0$ such that the Fisher information of a solution $w$ of~(\ref{cormoran1}) satisfies for $t\geq 0$
$$
\int |\partial_x \sqrt{w(t)}|^2 \,dx\leq \left( \int |\partial_x \sqrt{w(t=0)} |^2 \,dx\right) e^{-\mu L t}
$$
Indeed, a small computation gives
$$
\frac{d}{dt} \int |\partial_x \sqrt{w(t)}|^2 \,dx = L \int \left( \partial_x^4 w - \partial_x^2 \left( \frac{(\partial_x w)^2}{w} \right) \right)\left( 2 \frac{\partial_x^2 w}{w} - \left( \frac{\partial_x w}{w} \right)^2 \right) \,dx.
$$
The inequality (3.3) in~\cite{DGJ} gives a majorization of the above right-hand side by $\displaystyle - \mu L \int |\partial_x^3 \sqrt{w} |^2 \,dx$, for a constant $\mu>0$. This leads to the differential inequality
$$
\frac{d}{dt} \int |\partial_x \sqrt{w(t)}|^2 \,dx \leq -\mu L \int |\partial_x^3 \sqrt{w(t)}|^2 \,dx,
$$
from which the desired result follows by Poincar\'e's inequality.

\subsubsection{The case $x \in \mathbb{R}$, $K=0$} Still under the assumption that $\int w =1$, it was established in~\cite{MMS} that the solution $w$ of \eqref{cormoran1} converges to a Gaussian:
\begin{align}\label{Gauss-limit}
w \sim \frac{1}{\sqrt{\pi} t^{1/4}} e^{-\frac{x^2}{\sqrt{t}}} \quad \mbox{as $t \rightarrow \infty$.}
\end{align}

\subsubsection{The case $x \in \mathbb{R}$, $L=0$}
If $L=0$, $v$ is simply a solution of the Airy equation, for which the asymptotics are
$$
v(t,x) \sim \frac{1}{t^{1/3}} \frak{Re} \left[ \widetilde{\operatorname{Ai}} \left( \frac{x}{(-3K t)^{1/3}} \right) F \left( \frac{x}{t} \right) \right]\quad \mbox{as $t \rightarrow \infty$},
$$
where $F$ is complex-valued and can be expressed in terms of the Fourier transform of the initial data, while the modified Airy function $\widetilde{\operatorname{Ai}}$ is given by
$$
\widetilde{\operatorname{Ai}}(z) = \int_0^\infty e^{ix\xi+i \frac{\xi^3}{3}}\,d\xi
$$
(this is classical, see for instance~\cite{HN}, equation (2.3)).

\subsubsection{The case $x \in \mathbb{R}$, $K,L \neq 0$} This is the most interesting case, but it seems very difficult to analyze. It is argued heuristically in~\cite{Toom} that the invariant law for the random process that~\eqref{cormoran2} is supposed to model should be given by a rescaling of the Tracy-Widom distribution $F_1$. This prediction is then confirmed numerically. For the equation~\eqref{cormoran2}, it leads us to conjecture the following asymptotics:
$$
w(t,x) \sim \frac{2}{(6Kt)^{1/3}} F_1' \left( \frac{2x}{(6Kt)^{1/3}} \right) \qquad \mbox{as  $t\to \infty$}.
$$

\section{Invariant probability measure}

\label{sectioninvariant}

In this section, we study the invariant probability measure of the DLSS Markov process. We find more convenient to study the interacting particle process. The invariant measure depends only the ratio $\lambda = \lambda_+/\lambda_-$. Without loss of generality, we set 
$$
\lambda_-  =1  \qquad \mbox{and} \qquad \lambda_+ = \lambda. 
$$
We restrict our attention to the first $n$ particles $X(t) = (X_1 (t), \cdots, X_n(t))$. We set $x_0 = 0$ and consider $x = (x_1, \cdots, x_n) \in \N^n$ with $0  < x_1 < \cdots < x_n$. If $u (t,  x ) = P ( X (t) = x)$, we find
\begin{eqnarray}\label{galinettecendree}
\frac{d}{dt} u(t,x)& = & \sum_{i = 1 } ^n \sum_{ k = 0}^{n-i}  \lambda u(t, x - e_{i,k}) \IND_{x_{i-1} \ne x_{i} - 1}  \prod_{\ell = 1}^{k} \IND_{x_{i+\ell} = x_i +\ell}   \\
 & &\quad + \quad \sum_{i = 1 } ^n \sum_{ k = 1}^{x_{i+1} - x_i -1} u(t, x +  k e_{i}) - \lambda n u(t,x) - (x_n - n) u(t,x),
\nonumber
\end{eqnarray}  
where $x_0 = 0$ and the vectors $e_{i,k}$ and $e_i = e_{i,0}$ are defined by $e_{i,k} (j) = \IND ( i \leq j \leq i + k)$. The first double sum corresponds to a move on the right of the $i$-th particle which may have pushed its right neighbors, the second sum is move on the left of the $i$-th particle. If $X(t)$ is stationary, i.e. $u(t,x) = u(x) = \dP (X = x)$ where $\dP$ is the invariant measure, we obtain the system of equations 
\begin{align}\label{galinettecendree2}
& \left(\lambda n + (x_n - n) \right)  u(x)  \\
 &\quad  = \quad  \sum_{i = 1 } ^n \left( \sum_{ k = 0}^{n-i}  \lambda u( x - e_{i,k}) \IND_{x_{i-1} \ne x_{i} - 1}  \prod_{\ell = 1}^{k} \IND_{x_{i+\ell} = x_i +\ell} \;  + \sum_{ k = 1}^{x_{i+1} - x_i -1} u( x +  k e_{i}) \right).
\nonumber
\end{align}  

The function $u = u_{n}$ depends implicitly on the total number of particles.   However, recall that the restriction property of the DLSS process implies that 
\begin{equation}\label{restriction}
u_n( x ) = \sum_{y = x_n +1 }^\infty u_{n+1}(( x, y)).  
\end{equation}
Hence with a slight abuse of notation, we will remove the explicit dependency in $n$ and set for $x \in \N^k$, $u(x) = u_k (x)$.

It is easy to solve \eqref{galinettecendree2} in the case $n=1$. It corresponds to the stationary distribution of the first $+$ in the DLSS process. For $x \in \dN$,  \eqref{galinettecendree2} reads
$$
(\lambda + x - 1) u(x) = \lambda u(x-1) \IND_{x \geq 2} + \sum_{k=1}^\infty u ( x + k).
$$
Since $ \sum_{k=1}^\infty u (k) =1$, we may rewrite the above equation as 
$$
(\lambda + x) u(x) =  \lambda u(x-1) \IND_{x \geq 2}  + 1 - \sum_{k=1}^{x-1} u(k). 
$$
We can solve this equation by recursion on $x \in \N$, we find 
$$
u (x) = \frac{ x \lambda^{x-1}}{ \gamma(x)}
$$
where $\gamma (x) = \prod_{k=1}^x (\lambda + k)$ is a gamma-type function. It is a direct consequence of the identity 
$$
\frac{ \lambda^x }{ \gamma (x) } = 1 - \sum_{k=1}^{x} \frac{ k \lambda^{k-1}}{\gamma (k)}.
$$
Note that the expression for $u_1(x)$ implies a faster than exponential tail. 

For $n=2$,  the computation is already difficult. For integers $1 \leq x < y$,  using \eqref{restriction}, we find similarly
\begin{eqnarray}
 (\lambda + y - 1) u(x,y) & = &\lambda u(x-1,y) \IND_{x \geq 2} + \lambda u(x-1,y-1) \IND_{y = x+1} \IND_{x \geq 2} \nonumber  \\
 &  & \quad  +  \; \lambda u(x,y-1) \IND_{ y \ne x + 1}  + \sum_{k=1}^{ y - x -1} u ( x + k, y) + u (x) - \sum_{k=1}^{ y - x - 1 } u ( x , y - k ) .\label{galinettecendree3}
\end{eqnarray}
 We see from this expression that $u(x,y)$ could in principle be computed by recursion on $y \geq 2$. Indeed, in \eqref{galinettecendree3} $u(x,y)$ is expressed in terms of $u(x)$  and $u(x',y')$, $1 \leq x' < y ' \leq y-1$ (this remark extends to any number of particles $n \geq 1$).  The computation of the first terms gives
$$
u(1,2) = \frac{1}{(\lambda +1)^2} , \; u(2,3) = \frac{ \lambda ( 5\lambda + 4)}{(\lambda+1)^2 (\lambda^2 + 3 \lambda + 4)} , \; u(1,3) = \frac{ \lambda ( 2 \lambda^2 + 11 \lambda  + 8)}{(\lambda+1)^2(\lambda + 2) (\lambda^2 + 3 \lambda + 4)}.
$$
We have however not been able to find a closed-form formula for all $x < y$. 

\bigskip

\section{Moderate-time numerical simulation}\label{section:numerics}

Here we develop a stable numerical scheme to approximate solutions of the initial-value problem of \eqref{eqw}.  We investigate the limiting form of solutions for moderate times.  We write the equation \eqref{eqv} for $ v = \sqrt{w}$ assuming $w > 0$ so that
\begin{align}
v_t &= F(v) = F(\sqrt{w}),\notag\\
w_t &= 2 \sqrt{w} F(\sqrt{w}) = \frac{8}{3} \mu C \sqrt{w}\partial_x^3 \sqrt{w} + 2\epsilon (2 C^2 - 2 C) \left( \sqrt{w} \partial_x^4 \sqrt{w} - (\partial_x^2 \sqrt{w})^2 \right).\label{no-div}
\end{align}
In this way of writing the equation, we have no division operations.  While some issues could persist from performing the square-root, the smoothness and exponential decay of the solution make this operation accurate.  We employ a standard technique to compute solutions of \eqref{no-div}.  Let $\ell > 0$ and consider \eqref{no-div} on the periodic interval $(-\ell,\ell]$ with initial data $w_0(x)$.  We choose $w_0(x)$ to be an exponentially decaying function defined on $\mathbb R$ and $\ell>0$ sufficiently large so that $|w_0(x)|$ is less than, say, $10^{-16}$ outside $(-\ell,\ell]$.  We also have to choose $\ell$ sufficiently large so that the approximate solution remains zero (or approximately zero) near the boundary points $\pm\ell$ for the largest $t$ used in the computation.

From here, the problem fits into the classical theory for the numerical solution of time-dependent problems, see \cite[Section~9.6]{Boyd}.  One uses the pseudospectral differentiation operator $\mathcal D_{n,\ell}$ to approximate the derivatives in the right-hand side.  More precisely, the operator is described by the following schematic  (FFT stands for the Fast Fourier transform):
\begin{align*}
\begin{array}{c|c}
f:(-\ell, \ell] \rightarrow \mathbb C ~\overset{2^n~\text{sample points}}{\longrightarrow}~ (f(x_{1,\ell}), f(x_{2,\ell}), \ldots, f(x_{2^n,\ell}))^\top &  x_{m,\ell} = -\ell + 2\ell \frac{m+1}{2^n+1},\\ \\
 (f(x_{1,\ell}), f(x_{2,\ell}), \ldots, f(x_{2^n,\ell}))^\top ~\overset{\text{FFT}}{\longrightarrow}~ (\hat f_{-2^{n-1}+1},\hat f_{-2^{n-1}+2}, \ldots, \hat f_{2^{n-1}})^\top & f(x) \approx \sum \hat f_k e^{i k \pi/\ell x},\\ \\
\hat f_k ~\overset{\text{differentiate}}{\longrightarrow}~\tilde f_k := \frac{i k \pi}{\ell} \hat f_k & f'(x) \approx \sum \tilde f_k e^{i k \pi/\ell x}\\ \\
(\tilde f_{-2^{n-1}+1},\tilde f_{-2^{n-1}+2}, \ldots, \tilde f_{2^{n-1}}) ~\overset{\text{inverse FFT}}{\longrightarrow}~ D_{n,\ell}(f(x_{1,\ell}), f(x_{2,\ell}), \ldots, f(x_{2^n,\ell}))^{\top}
\end{array}
\end{align*}
The end result of this is that if $f$ is sufficiently smooth (and periodic) then
\begin{align*}
D_{n,\ell}(f(x_{1,\ell}), f(x_{2,\ell}), \ldots, f(x_{2^n,\ell}))^{\top} \approx (f'(x_{1,\ell}), f'(x_{2,\ell}), \ldots, f'(x_{2^n,\ell}))^\top,
\end{align*}
is a good approximation.  This allows us to accurately approximate the right-hand side of \eqref{no-div}.  We use the fourth-order Runge--Kutta method to time step the solution.  Explicitly, given a time step $h > 0$, the method is for $m \geq 0$
\begin{align*}
\mathbf w(t_0) &= (w_0(x_{1,\ell}), w_0(x_{2,\ell}), \ldots, w_0(x_{2^n,\ell}))^{\top},\\
\mathbf w(t_{m+1}) &= \mathbf w(t_{m}) + \frac{h}{6} \left( \mathbf k_1 +  \mathbf k_2 + \mathbf k_3 + \mathbf k_4\right),\\
t_0 &= 0, ~~ t_{m+1} = t_m +h,\\
\mathbf k_1 &= F_{m,\ell}(\mathbf w(t_{m})),\\
\mathbf k_2 &= F_{m,\ell}(\mathbf w(t_{m}) + \frac{h}{2} \mathbf k_1),\\
\mathbf k_3 &= F_{m,\ell}(\mathbf w(t_{m}) + \frac{h}{2} \mathbf k_2),\\
\mathbf k_4 &= F_{m,\ell}(\mathbf w(t_{m}) + h \mathbf k_3),\\
F_{n,\ell}(\mathbf w) &= \frac{8}{3} \mu C \sqrt{\mathbf w} \mathcal D^3_{n,\ell} \sqrt{\mathbf w} + 2\epsilon (2 C^2 - 2 C) \left( \sqrt{\mathbf w} \mathcal D^4_{n,\ell} \sqrt{\mathbf w} - (\mathcal D^2_{n,\ell} \sqrt{\mathbf w})^2 \right).
\end{align*}

We highlight a numerical complication.  Often, when time-stepping a time-evolution PDE with a high-order linear term, one wants to treat the linear term explicitly.  This is the so-called method of exponential integrators, see \cite{Kassam2005}, for example. But, it is clear that the linear terms in \eqref{eqw}, treated alone, will cause the solution to vanish, making it impossible to apply the nonlinear terms.  Thus, there must be a close interplay between the linear and nonlinear terms in \eqref{no-div} and we cannot use exponential integrators.  This forces a small time step. We are still able to perform simulations for moderate times but at a much higher computational cost.

\subsection{The Gaussian limit}

\begin{figure}[tbp]
\subfigure[]{\includegraphics[width=.48\linewidth]{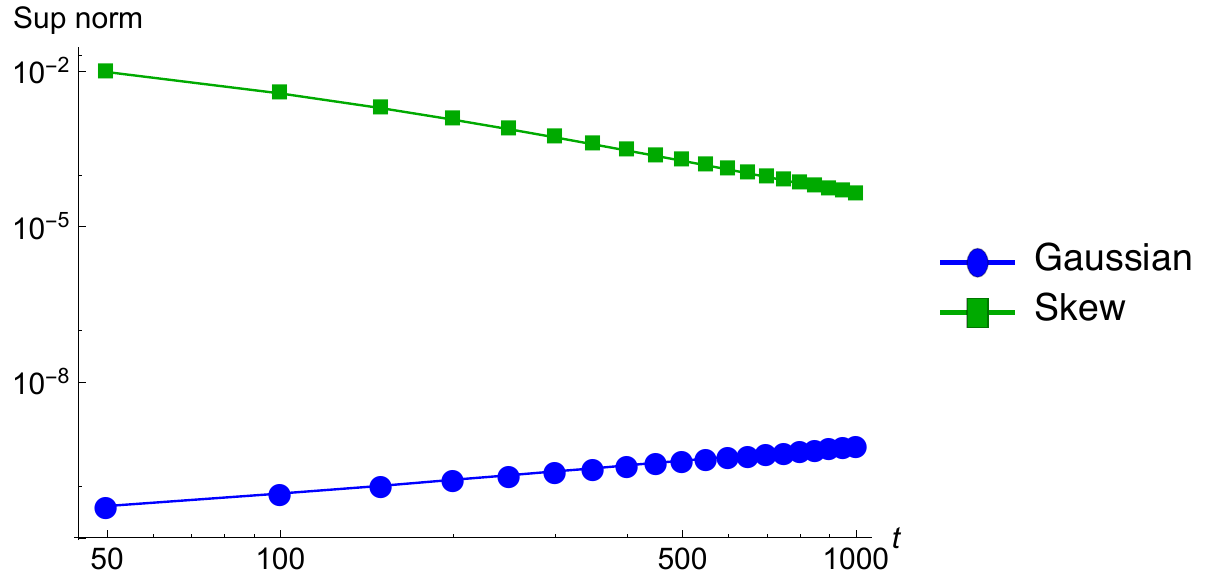} \label{fig:Gaussian}}
\subfigure[]{\includegraphics[width=.48\linewidth]{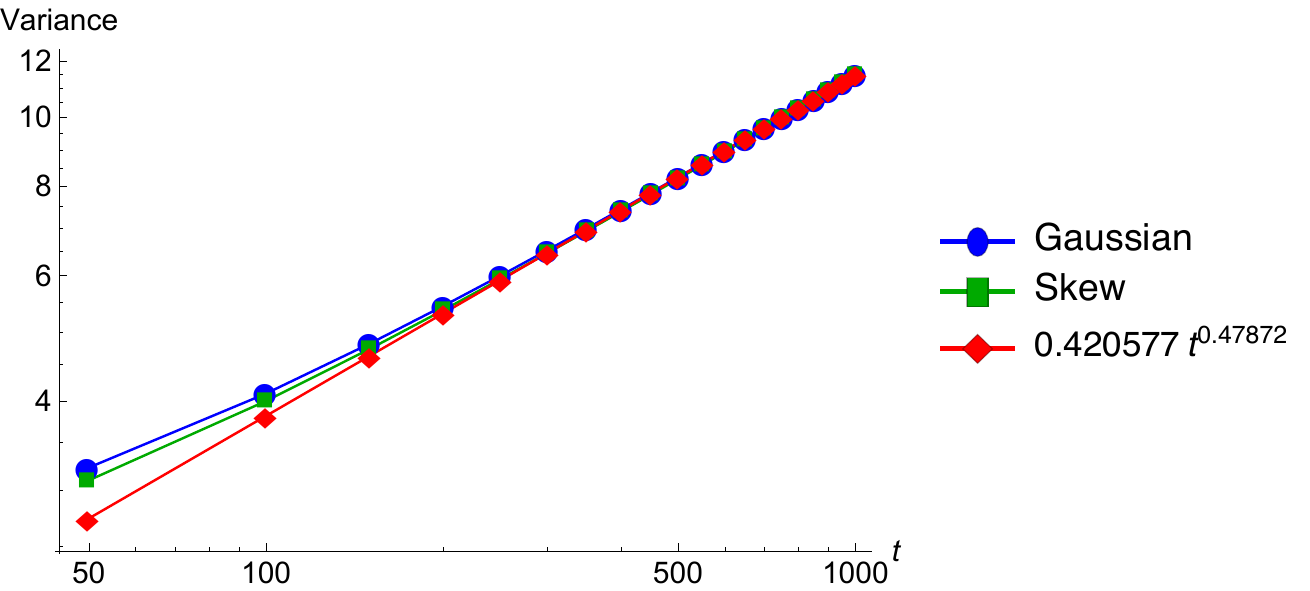} \label{fig:Gaussian-var}}
\caption{(a) An estimate of the difference $\sup |\bar w(t_m,x)-g(x)|$ for as $t_m$ increases for both the skew initial condition $w_0 = e^{-x - e^{-x}}$ and the Gaussian initial condition $w_0 = \frac{1}{(2\pi)^{1/4}}e^{-x^2/4}$.  The difference stays small with for the Gaussian initial data and the difference decreases in time for the skew initial data. (b) The comparison of the scaling of the variance $c_m-b_m^2$ as a function of $t$.  The diamonds correspond to the least-squares fit with equation $.420577 t^{0.47872}$ which is close to expected $t^{1/2}$ scaling.  We note that the least-squares fit is only performed for $t > 500$.}
\end{figure}

If $\mu = 0$ then the solution of \eqref{no-div} should limit to the Gaussian similarity solution as was shown in \cite{MMS}.  To test our numerical scheme on this we do the following.
\begin{itemize}
\item Set $w_0(x) = \frac{1}{(2\pi)^{1/4}}e^{-x^2/4}$, \emph{i.e.} we start with a non-standard Gaussian density.
\item At each $t_m$, approximate $a_m =\int w(t_m,x) dx$, $b_m = a_m^{-1}\int x w_0(t_m,x) dx$ and \\$c_m = a_m^{-1} \int x^2 w_0(t_m,x) dx$ with the trapezoidal rule.
\item Define $\bar w(t_m,x) = \frac{c_m^{1/2}}{a_m} w(t_m,c_m^{1/2}x + b_m)$.  This is a probability density with mean zero and variance one.
\item We monitor how close $\bar w(t_m,x)$ is to a standard Gaussian density $g(x) := (2 \pi)^{-1/2} e^{-x^2/2}$ with an estimate of the supremum norm. 
\end{itemize}
To be precise, we use $\epsilon = .1$, $C = .2$, $n = 10$ and $h = 0.0025$. See Figure~\ref{fig:Gaussian} for a demonstration of convergence to the Gaussian limit.  We also run this same calculation with $w_0(x) = e^{-x - e^{-x}}$ and show the results in Figure~\ref{fig:Gaussian}.  From \eqref{Gauss-limit}, the variance $c_m - b_m^2$ should scale like $t^{1/2}$ and we confirm this in Figure~\ref{fig:Gaussian-var}.

\subsection{The biased case}

When $\mu \neq 0$, if the conjecture made in \cite{Toom} is correct and carries through the formal derivation above,  we should see $\bar w(t_m,x)$ converge to the density for the Tracy--Widom ($\beta = 1$) GOE distribution after it is normalized to mean zero and variance one and possibly reflected across $x = 0$ (due the sign of $\mu$).  We call this normalized density $f_1(x)$.  It is easily computed once one can compute the Hastings--McLeod solution of the Painlev\'e II equation, see \cite{TrogdonSORMT}.  We perform the same computations as in the previous section but now with multiple choices for initial data to examine the convergence deeper.  We choose the following functions for initial data
\begin{align*}
w_0(x) &= \frac{1}{(2 \pi)^{1/4}} e^{-x^2/4},  ~~ (\text{``Gaussian"})\\
w_0(x) &= \frac{1}{(2 \pi)^{1/4}} e^{-(x+2)^2/4} + \frac{1}{(2 \pi)^{1/4}} e^{-(x-2)^2/4}, ~~ (\text{``Mixed Gaussians"})\\
w_0(x) &= f_1(x),  ~~ (\text{``Tracy--Widom"}).
\end{align*}

\begin{figure}[tbp]
\subfigure[]{\includegraphics[width=.48\linewidth]{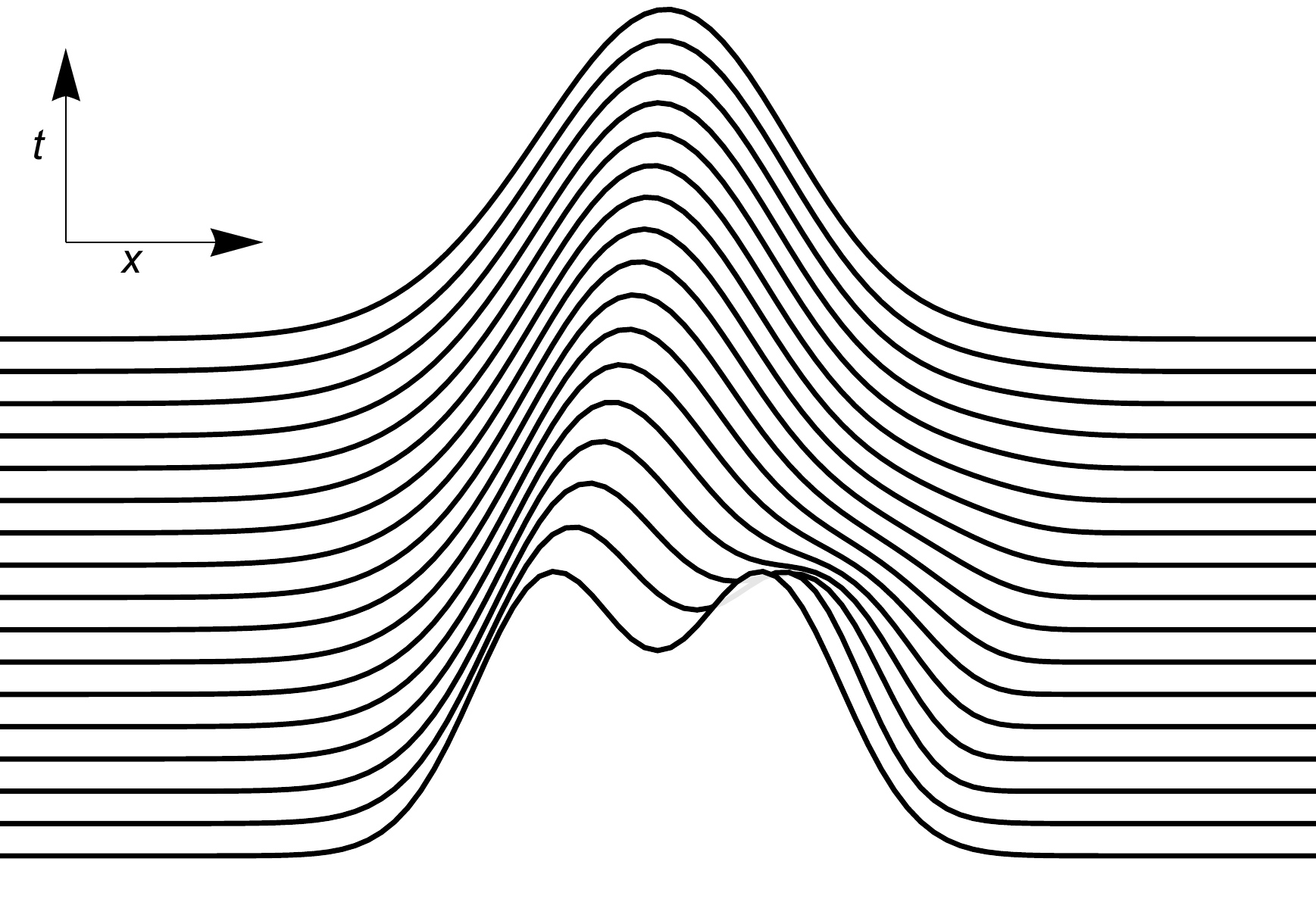}}
\subfigure[]{\includegraphics[width=.48\linewidth]{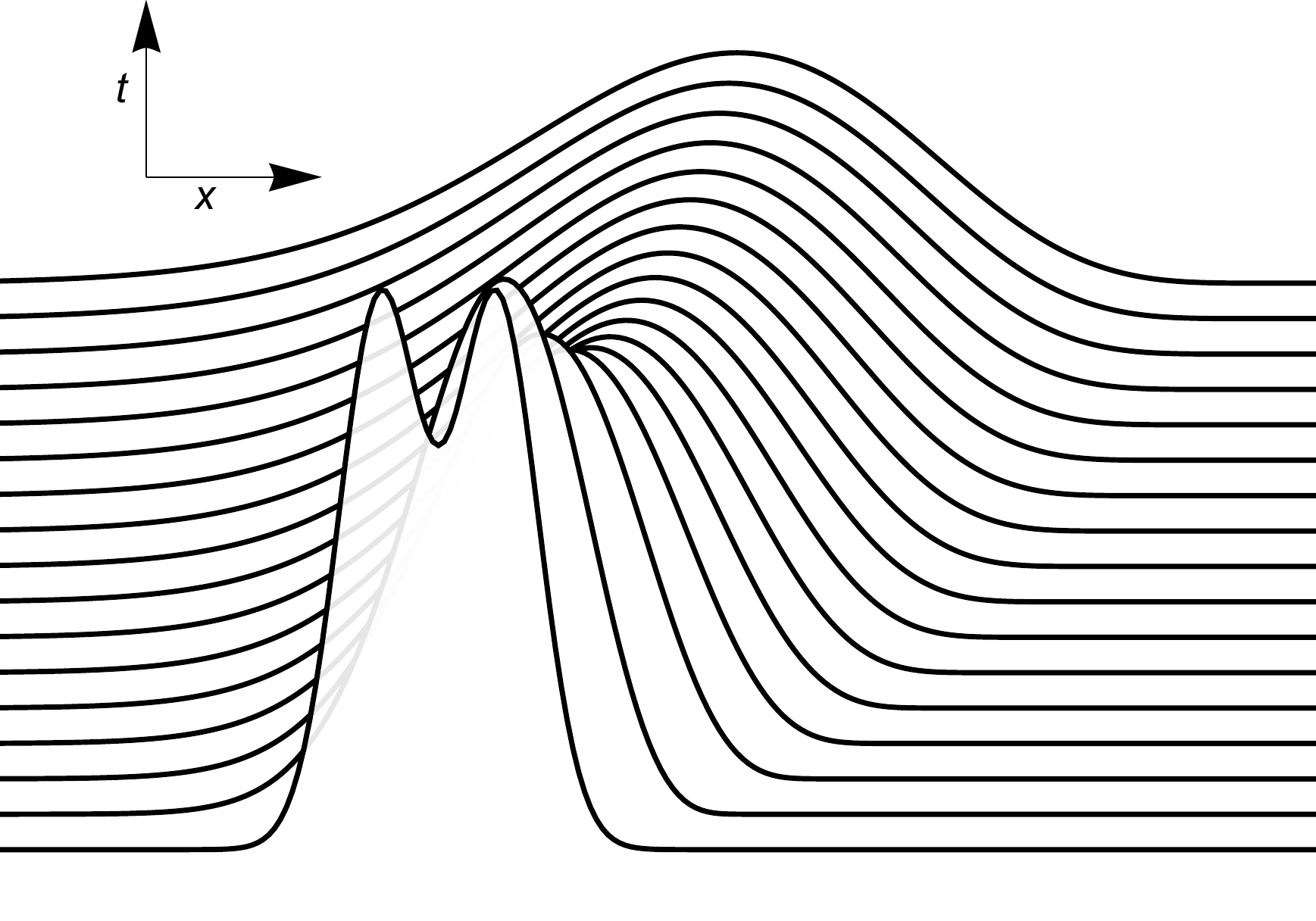} }
\caption{\label{fig:evolve} (a) The evolution of the approximation of $\bar w(t_m,x)$ as $t_m$ increases from $t_0 = 0$ to $t_m = 20$ with the Mixed Gaussians initial data.  (b) The evolution of the approximation of $w(t_m,x)$ as $t_m$ increases from $t_0 = 0$ to $t_m = 900$ with the Mixed Gaussians initial data.}
\end{figure}

We use $\mu =1$, $\epsilon = .1$, $C = .2$, $n = 10$ and $h = 0.0025$.  In all cases we consider, our numerical method preserves the $L^1(\mathbb R)$ norm of the solution to within $10^{-10}$.  It is approximately conserved and this is a consistency check on the numerical method. In Figure~\ref{fig:evolve} we plot the evolution of the Mixed Gaussians initial data under the flow.
\begin{figure}[tbp]
\begin{center}
\subfigure[]{\includegraphics[width=.4\linewidth]{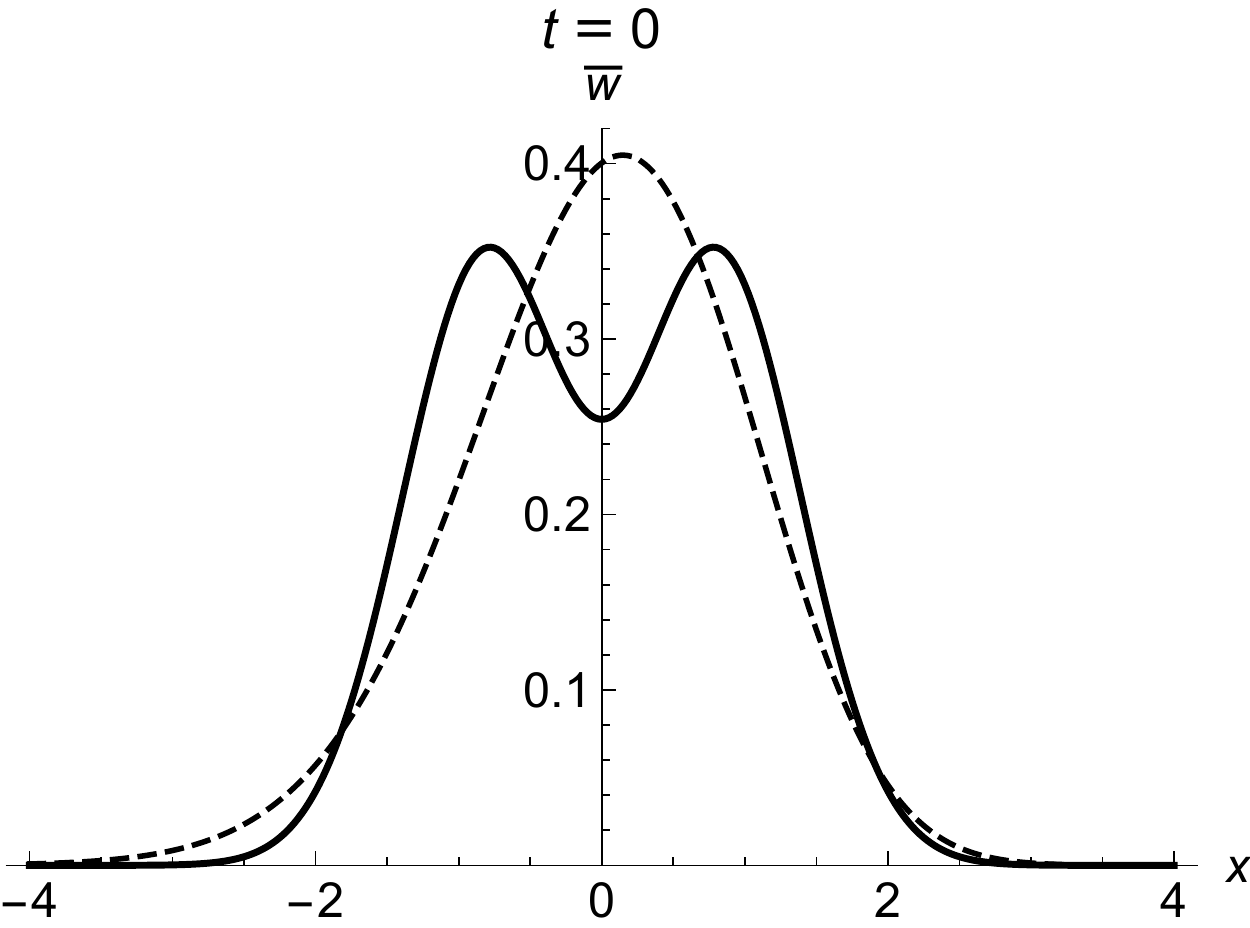}}
\subfigure[]{\includegraphics[width=.4\linewidth]{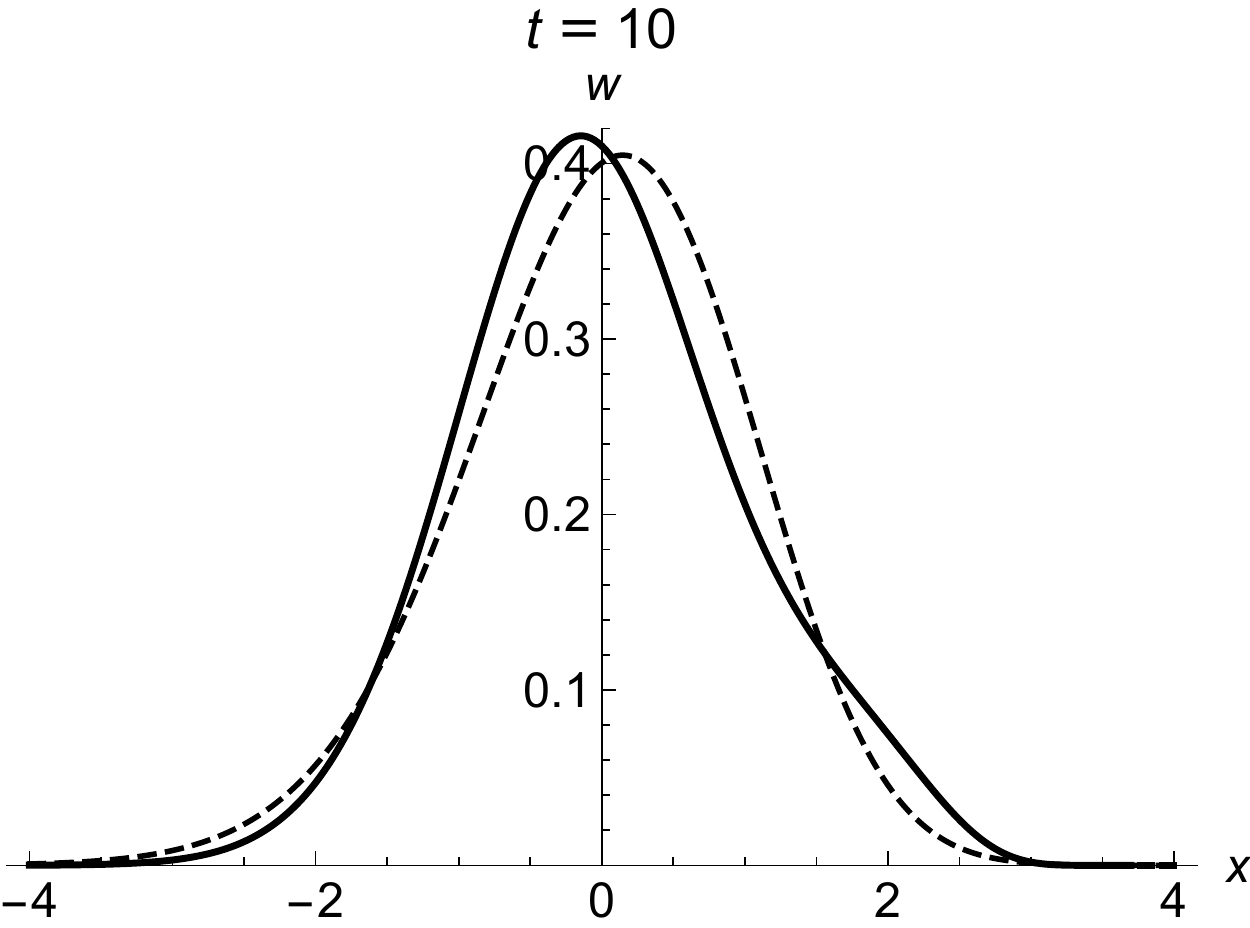}}
\subfigure[]{\includegraphics[width=.4\linewidth]{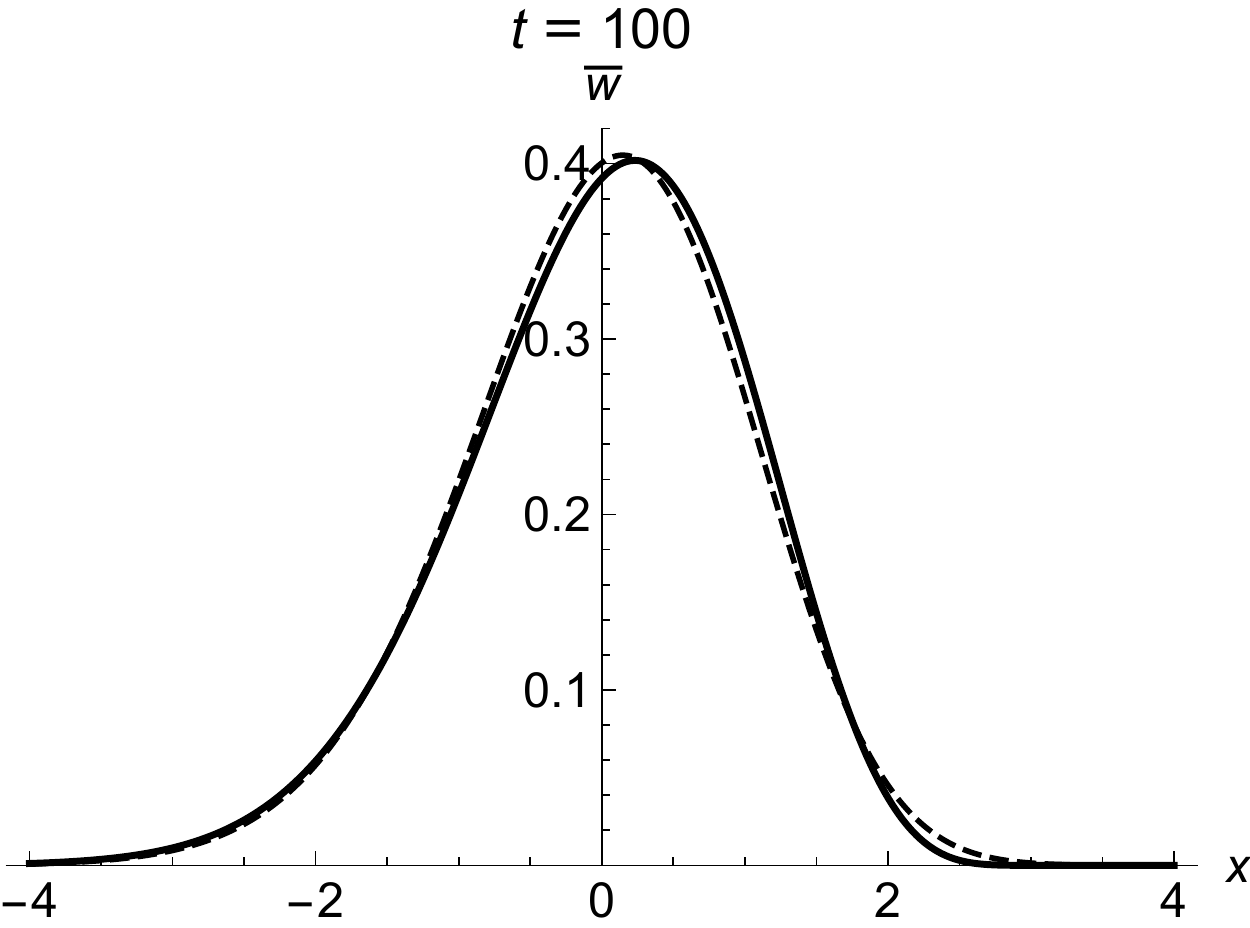}}
\subfigure[]{\includegraphics[width=.4\linewidth]{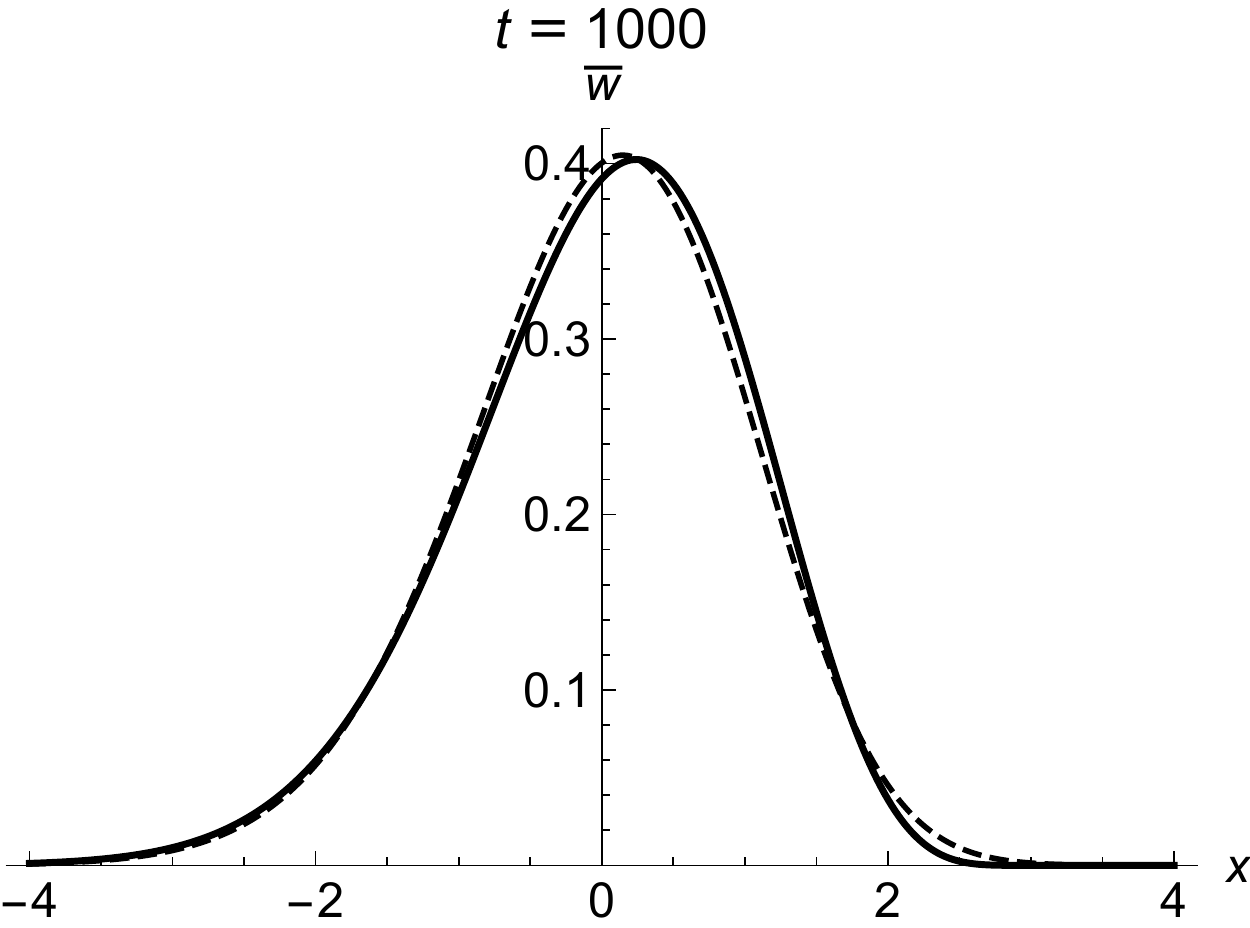}}
\end{center}

\caption{A view of the evolution of the approximation of $\bar w(t_m,x)$ (solid) in comparison to $f_1(x)$ (dashed). The limiting form of $\bar w(t_m,x)$ appears to be close to $f_1(x)$ but still distinct. (a) The scaled initial condition $\bar w(0,x)$. (b) The approximation of $\bar w(10,x)$. (c) The approximation of $\bar w(100,x)$. (d) The approximation of $\bar w(1000,x)$. }
\end{figure}
 In Figures~\ref{fig:twmean} and \ref{fig:twvar} we plot the mean $b_m$ and the variance $\sigma_m^2 = c_m-b_m^2$ of the solution as a function of $t$ for each of the choices of initial data on log-log axes.  Performing a least-squares fit we conjecture that $b_m \sim t^{1/3}$ and $\sigma_m^2 \sim t^{2/3}$ for large $t$.  To see that $b_m \sim t^{1/3}$ we consider the Fisher information $I_m$ in Figure~\ref{fig:fisher} which is, as discussed above, the time derivative of the mean.  It is clear here that $f_m \sim t^{-2/3}$ implying that $b_m \sim t^{1/3}$.  The discrepancy in the exponent of our least-squares fit in Figure~\ref{fig:twmean} appears to be due to $\mathcal O(1)$ or $\mathcal O(\log t)$ terms that arise from integrating the Fisher information. This all means that we have a limiting form of
\begin{align}\label{w-limit}
w(t,x) \sim \frac{1}{t^{1/3}} f \left( \frac{x}{t^{1/3}} \right).
\end{align}
Finally, to see that $f$, in our experiments, exists empirically but is distinct from $f_1$, we plot estimates of the difference $\sup_{\mathbb R}|f_1(x) -\bar w(t_m,x)|$ for a series of times for each initial condition in Figures~\ref{fig:twsup} and \ref{fig:twsupdiff}.  From this it appears that an $f$ in \eqref{w-limit} exists but is differs from $f_1(x)$ by approximately $2 \times 10^{-2}$.  At this point, these results are intriguing but we cannot claim to refute or substantiate the conjecture in \cite{Toom}.  An additional intriguing detail is that the computations in \cite[Figure~3]{Toom} appear to have densities that differ from a scaled Tracy--Widom GOE by $3 \times 10^{-2}$.  Without accounting for the normalization of the mean and the variance, our computations cannot be compared qualitatively with these other than to say that the errors are on the same order of magnitude and are therefore consistent.
\begin{figure}[p]
\subfigure[]{\includegraphics[width=.48\linewidth]{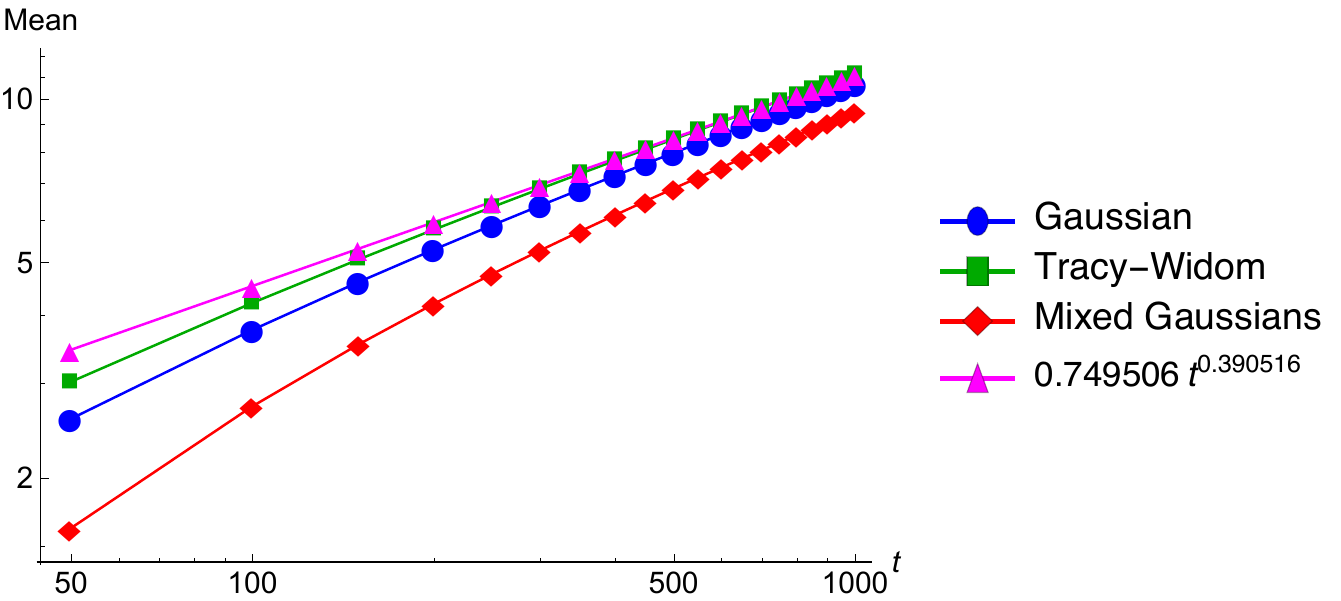}\label{fig:twmean}}
\subfigure[]{\includegraphics[width=.48\linewidth]{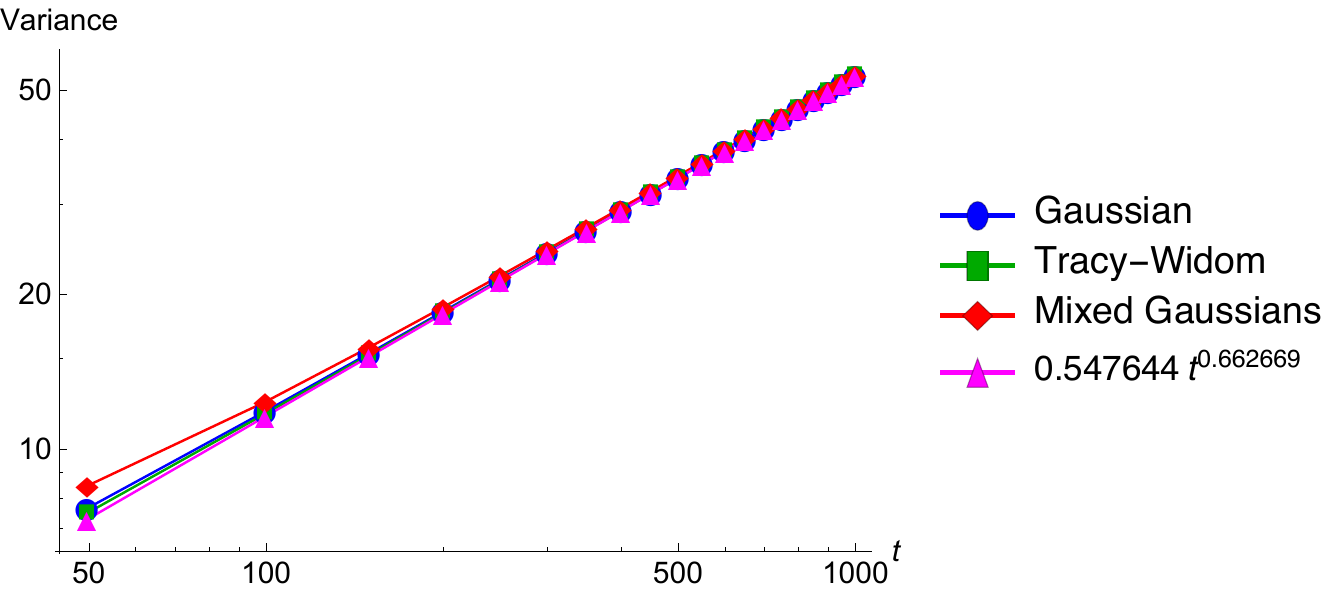} \label{fig:twvar}}
\subfigure[]{\includegraphics[width=.48\linewidth]{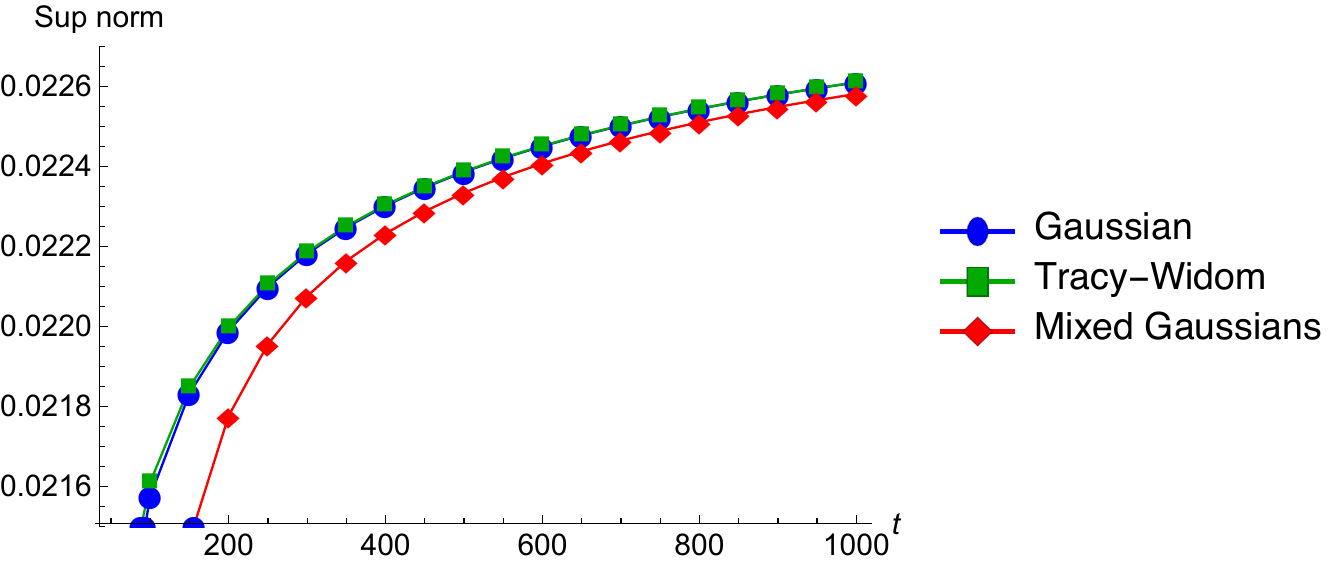} \label{fig:twsup}}
\subfigure[]{\includegraphics[width=.48\linewidth]{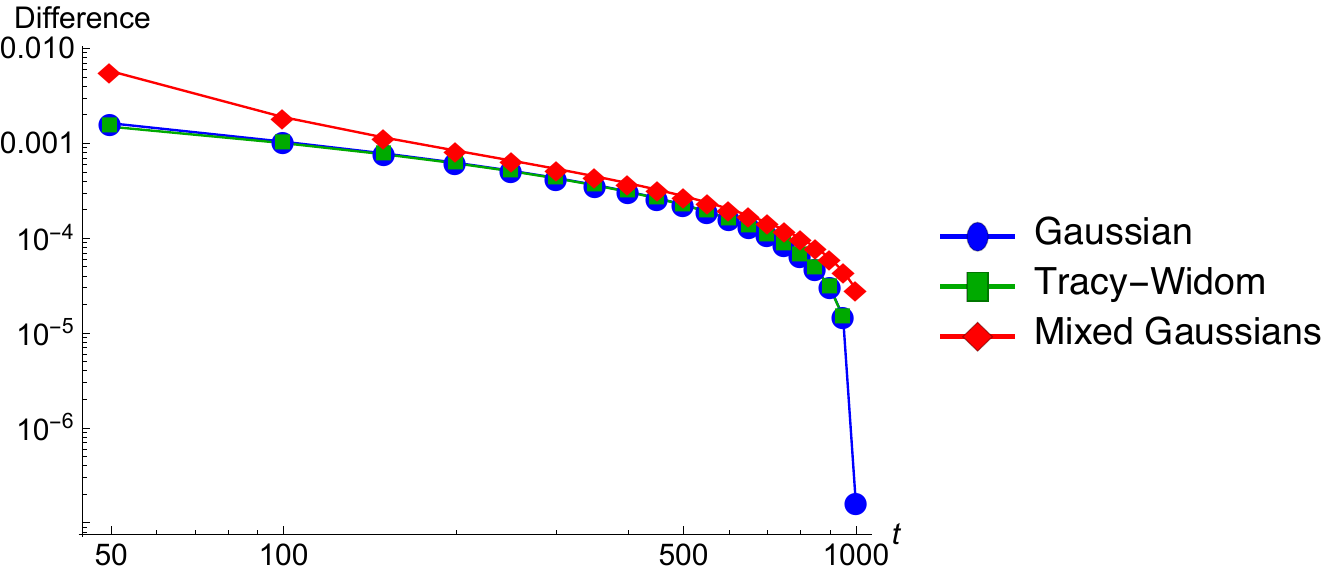}\label{fig:twsupdiff} }
\subfigure[]{\includegraphics[width=.48\linewidth]{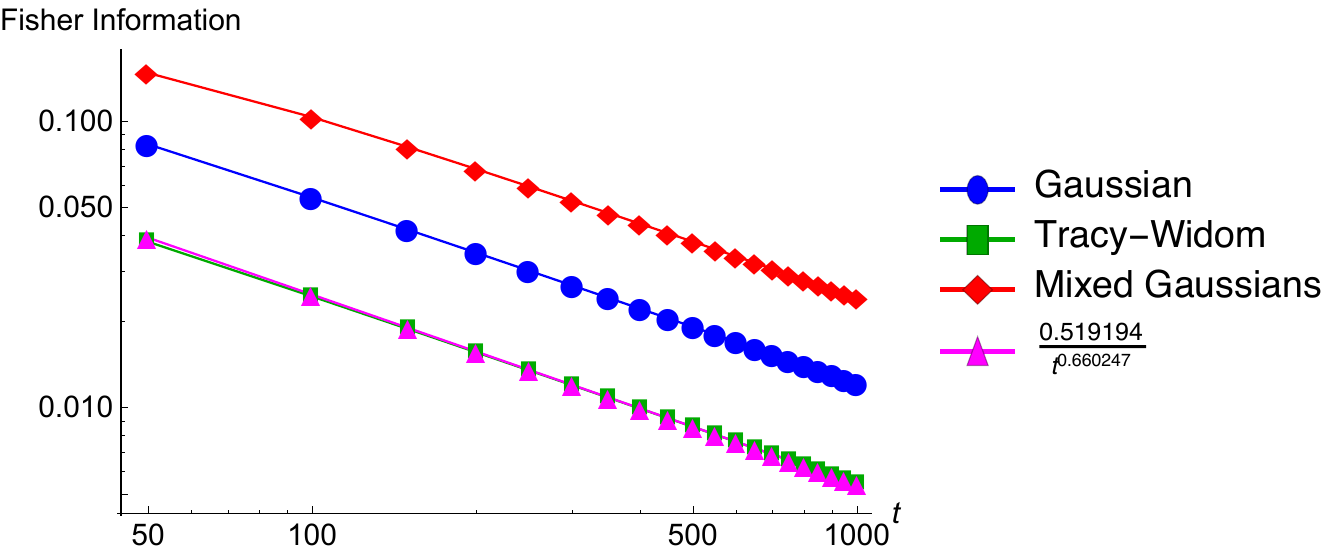}\label{fig:fisher} }
\caption{ (a) We plot the evolution of the mean $b_m$ for each choice of initial data.  A least-squares estimate for $t > 950$ gives $b_m \sim t^{0.390}$. When considering the Fisher information in plot (e), we conjecture that the true mean scales like $t^{1/3}$.  (b) We plot the evolution of the variance $\sigma_m^2= c_m-b_m^2$ for each choice of initial data.  A least-squares estimate for $t > 950$ gives $\sigma_m^2 \sim t^{0.66}$.  (c) The evolution of the approximation of $\sup |\bar w(t_m,x) - f_1(x)|$ for the Gaussian, Mixed Gaussians and Tracy--Widom initial data. There is a clear limiting value.  (d) Define $c =  \sup |\bar w(1000,x) - f_1(x)|$ when $w_0(x)$ is the Tracy--Widom initial data.  In this plot we examine $|c-\sup |\bar w(t_m,x) - f_1(x)||$ for each choice of initial data.  It is clear that they all have the same limiting value. (e) The evolution of the Fisher information $I_m$ for $w(t_m,x)$.  A least-squares fit for $t > 950$ gives $I_m \sim t^{-2/3}$ which implies that $b_m \sim t^{1/3}$. }
\end{figure}

If the true limiting state of the system, after normalization, is $f_1(x)$ the following reasons could explain our discrepancy:
\begin{itemize}
\item The periodic approximation excites an instability that acts in $\mathcal O(1)$ time and is sufficient to eliminate convergence.  This seems unlikely because as $\ell$ and $n$ are increased with $h$ being decreased the solution does not appear to close in on $f_1(x)$.
\item The expansion in $\epsilon$ must be carried out to higher orders to achieve greater accuracy.
\end{itemize}

\vspace{.2in}

\noindent \textbf{Acknowledgments.} The authors are grateful to Percy Deift for very helpful discussions while this article was being prepared.

\clearpage
\bibliographystyle{abbrv}

\bibliography{mat}

\end{document}